\documentclass[12pt,preprint]{aastex}
\shorttitle{A multiwavelength study of Galactic HII region Sh2-294}
\shortauthors{Samal et al.}
\begin{document}

\title{A multiwavelength study of Galactic HII region Sh2-294}

\author{M.R. Samal\altaffilmark{1}, 
A.K. Pandey\altaffilmark{2,3,1},
D.K. Ojha\altaffilmark{3}, 
S.K. Ghosh\altaffilmark{3},  
V.K. Kulkarni\altaffilmark{4}, 
B.C. Bhatt\altaffilmark{5}} 

\email{manash@aries.ernet.in, ojha@tifr.res.in}

\altaffiltext{1}{Aryabhatta Research Institute of Observational Sciences, Nainital, 263 129, India}
\altaffiltext{2}{Institute of Astronomy, National Central University, Chung-Li 32054, Taiwan}
\altaffiltext{3}{Tata Institute of Fundamental Research, Mumbai (Bombay), 400 005, India}
\altaffiltext{4}{National Center for Radio Astrophysics, Post Bag 3, Ganeshkhind, Pune 411007, India}
\altaffiltext{5}{Indian Institute of Astrophysics, Koramangala, Bangalore 560 034, India}

\begin{abstract}
We present the observational results  of Galactic HII region Sh2-294,
using optical  photometry, narrow-band imaging and  radio continuum mapping 
at 1280  MHz, together  with archival  data  from 2MASS,  MSX and IRAS 
surveys.  The stellar  surface density  profile based on 2MASS data 
indicates that  the radius of  the cluster  associated with the Sh2-294 region is  
 $\sim 2.3^{\prime}$. We found  an anomalous  reddening law  for  
the dust  inside the  cluster
region and the  ratio of total-to-selective extinction ($R_V$) is
found to be  3.8 $ \pm$  0.1. We estimate the minimum reddening 
$E (B-V)$ = 1.35 mag and distance of 4.8 $\pm$ 0.2  kpc to the region from
optical color-color and  color-magnitude diagrams. We identified 
the ionizing source of the HII region, and spectral type estimates based 
on different methods are  consistent with a star  of spectral type $\sim$ B0 V. 
The 2MASS $JHK_{\rm s}$ images reveal a  partially embedded cluster associated with 
the ionizing source along  with a small cluster towards the eastern border
of  Sh2-294. The ionization 
front  seen along  the  direction of small cluster in radio continuum 
and H$\alpha$ images, might be due to
the  interaction of ionizing sources  with the  nearby  molecular cloud.
We found an  arc  shaped 
diffuse molecular hydrogen emission at  2.12 $\mu$m and a half ring of
MSX  dust emission  which surrounds  the  ionized gas  in  the direction  of
the ionization front.
The $HIRES$ processed $IRAS$
maps show two clumpy structures of high optical depth ($\tau_{100}$) and low color temperature
($T(60/100)$) at the eastern border of the nebula.
Self consistent radiative transfer model of mid- to far-infrared continuum
emission detected near small cluster is in good agreement with the observed
spectral energy distribution of a B1.5 ZAMS star.
The morphological correlation between the ionised and  molecular gas,
along with probable time scale involved between the ionising star, 
evolution of HII region and  small cluster, indicates that the  
star-formation activity observed  at the  border is probably
triggered by  the expansion  of HII region.
\end{abstract}



\section{Introduction}

Massive OB star formation takes place in the  dense core of molecular
cloud and are associated with dense stellar clusters. Their pre-main-sequence
(PMS) life scales are much shorter  compared to low mass stars.  The stellar
winds and  radiation emitted from  these stars eventually  disrupt the
molecular cloud, dissociate and ionize the gas and form the HII regions.
The  radiation from  such stars  can modify  the interstellar
grains  present  in  their  immediate  vicinity, hence  affect  the
interstellar  extinction law (cf. Chini \& Wargau 1994; 
Pandey et al. 2000). 
Thus, change  in the  interstellar extinction
law  affects the  conclusion, such  as photometric  distances  and age
determination  of  the   clusters  from  comparison  with  theoretical
isochrones.  All  these short-comings therefore make it difficult to
understand  the  formation and early phase evolution of such objects.
The  interaction  of  large
radiation  fields of massive  stars with  dense molecular  gas produce
photodissociation regions  (PDRs),  which is  an  important tool  to
understand  the star formation,  as  evolution of  nearby  dense cloud  is
primarily governed by these  interactions.  The interaction of shock 
generated by the massive
stars  may  also be responsible  for  triggering  further star  formation
within the molecular cloud in  which they are born (Elmegreen \& Lada 1977).

In    this   paper,    we   have    studied   Sh2-294/RCW3   region
($\alpha_{2000}=07^{h}16^{m}34.5^{s}$,
$\delta_{2000}=-09^{\circ}26^{\prime}38^{\prime\prime}$)   which   was
identified as an HII region (Sharpless 1959) and is part of Monoceros
cloud complex.  In the optical, it  appears like a  butterfly with two
wings  separated by dark-lane,  roughly east-west.   In the  center of
dark  lane, a B0.5  star is  detected (Moffat,  Jackson, \& Fitzgerald 1979). 
Non-detection of  H$_2$O maser in the search by Henkel et al. (1986) for the 
Sh2-294 region indicates that  the region is an evolved HII  region. The
radio  map obtained  by Felli et al. (1978) at  4.99 GHz,
shows good morphological agreement between the
nebulosities in optical
and radio upto a linear size  of $7^{\prime}$. The molecular cloud associated
with Sh2-294 region has been
observed in CO (Blair, Peters, \& Vanden Bout 1975; Blitz, Fich, \& Stark 1982).
The distance to
Sh2-294  in the literature ranges from 2.5 to  4.6 kpc as quoted by Felli \& Harten (1981).
The  Sh2-294  region is coincident  with an IRAS Point Source Catalog (PSC) 
source (IRAS 07141-0920) with an increasing spectrum from 12 to 100 $\mu$m.
Wouterloot \& Brand (1989) reported  kinematic distance to the IRAS 07141-0920
as 3.01 kpc and a luminosity of 3.6 $\times$ $10^3$ $L_\odot$ for this IRAS
source.

To continue  our  multiwavelength investigations  of massive star forming 
regions (cf. Ojha et al. 2004a; Ojha et al. 2004b; Ojha et al. 2004c; 
Tej et al.  2006), we have carried out a detailed study of Sh2-294 region 
as a useful example for understanding the massive star  formation
with the following layout of the paper.  In Sect. 2,
we  describe  our  observations  and  the  reduction  procedures.   In
Sect.  3, we  discuss other  available  datasets used  in the  present
study. Sect. 4 discusses the stellar component and deals with  the basic 
parameter determinations of the cluster associated to the massive star.
Sect.  5 describes the interstellar matter (ISM) component and 
the physical environment associated to  the  massive  star. Sect.  6 is
devoted  for the  discussion on  star  formation scenario  in Sh2-294
region and in Sect. 7 we present a summary of our results.

\section{ Observations and Data Reduction}

\subsection {Optical photometry}

$UBV(RI)_c$ CCD photometric  observations  were  performed  for  the
Sh2-294 region on 2005  December 26 and 2006  March 06, using
the 2K  $\times$ 2K  CCD system  at the f/13  Cassegrain focus  of the
104-cm  Sampurnanand telescope  (ST) of  ARIES, Nainital (India).   The 0.37
arcsec pixel$^{-1}$ plate scale  gives a field of view (FOV) of 
12$^{\prime}.6 \times 12^{\prime}.6$ on  the sky.  The read-out noise  for the system
was 5.3 $e^{-}$, while the gain  was 10 $e^{-}$ per ADU.  To improve
the signal-to-noise (S/N) ratio,  observations were made in 2 $\times$
2  pixel binning  mode. Several  bias frames  and twilight  flat field
exposures were taken during the observation.  We observed the standard
area SA101  (Landolt 1992) on 2006  March 06 several  times during the
night for  the purpose of the determination  of atmospheric extinction
coefficients  and   for  the   photometric  calibration  of   the  CCD
systems. The log of observations is  given in Table 1. The CCD images
were flattened using twilight sky  flats and bias subtraction was done
using  the IRAF\footnote{IRAF is distributed by the
National Optical Astronomy Observatories, which are operated by the
Association of Universities for Research in Astronomy, Inc., under contract
to the National Science Foundation.} data reduction  package. Then, for  
a given filter,
frames of  same exposure time were  combined into one,  to improve the
S/N enabling us to detect fainter stars.  The  photometric measurements of the stars were
performed  using DAOPHOT  II profile fitting  package of MIDAS.  The
stellar images were  well sampled and a variable  PSF was applied from
several  uncontaminated  stars  present   in  the  frames.
The photometric errors at brighter magnitude level (V $\leq 15$) 
are $\sim 0.01$ mag, which increase with V magnitude and become 
large ($\sim$ 0.1 mag) at V $\sim$ 22.5 mag.
The  image parameters (e.g. {\it chi \& sharpness}) and  errors provided  by 
DAOPHOT were  used to  reject poor
measurements and only the stars having error in magnitudes $\leq 0.1 $
are used in the further analyses.

\subsection {Optical spectroscopy}

To  study further,  a  spectroscopic observation
was made on 2006 March 16 for one of the sources, possibly 
responsible for the ionization of Sh2-294 region 
(cf. Moffat, Jackson, \& Fitzgerald 1979 and see \S 4.3). We used
the Himalaya  Faint Object Spectrograph Camera  (HFOSC) available with
the 2.01 m  Himalayan {\it Chandra} Telescope (HCT), operated  by the Indian
Institute of Astrophysics, Bangalore (India).
 The instrument is equipped with a  SITe 2K $\times$ 4K pixel CCD. The
 central 2K $\times$ 2K region, used for imaging, corresponds
 to   a  FOV  of   $\sim$   $10^{\prime}$  $\times$
 $10^{\prime}$, with a pixel size of 0.296 arcsec. 
The  slit spectra in  the wavelength range,  3700-6800 \AA,  were 
obtained  using low  resolution grism  \#7, with  a slit
 width  of $\sim$  2$^{\prime\prime}$. The  spectroscopic observations
 were  carried out under  good photometric  sky conditions.   Bias and
 flat  frames  were   obtained  at  the  beginning  and   end  of  the
 observations.  The  one-dimensional  spectra  were extracted  from  the
 bias-subtracted  and flat-field  corrected images  using  the optimal
 extraction  method. Wavelength  calibration of  the spectra  was done
 using FeAr lamp source.

\subsection {Optical and NIR Narrow-band imaging}

Optical CCD  narrow-band images of the nebula around Sh2-294 region were
obtained in  H$\alpha$ filter ($\lambda$ =  6565 \AA, $\Delta\lambda$ =
80 \AA)   and  its   nearby  continuum   filter   ($\lambda$=  6550 \AA,
$\Delta\lambda$  = 80 \AA) using 104-cm telescope at  ARIES,  on 2006
September 28. Narrow-band  near-infrared (NIR) observations
were  carried out in  the rotational-vibrational line of molecular 
hydrogen -- H$_2$(1-0)S1  ($\lambda$ = 2.122 $\mu$m,
$\Delta\lambda$ = 0.019   $\mu$m)  and   K continuum  (
$\lambda$ = 2.264  $\mu$m, $\Delta\lambda$  = 0.054 $\mu$m)  with  HCT on
2005 November 28 using the  NIR imager \mbox{(NIRCAM)}, which  is a 0.8 -  2.5 $\mu$m
camera with 512 $\times$ 512  HgCdTe array. For our observations, the NIRCAM was 
used in the {\it Camera-B} mode which has a FOV of $\sim$ 3$^\prime$.6
$\times$  3$^\prime$.6. We obtained several dithered (by $20^{\prime\prime}$) 
frames of the target in order to remove  bad pixels, cosmic rays and to
eliminate  the   presence  of  objects  with   extended  emission  for
construction of sky images.
All  images were  bias-subtracted and  flat-field  corrected in  the
standard manner using  IRAF package. The final images are produced by
subtraction of  continuum images after aligning and  PSF matching.

\subsection {Radio continuum observation}

In order to probe the ionized gas component,
radio continuum interferometric observation at  1280 MHz  was conducted  on 2006
December  8, using the  Giant Metrewave  Radio Telescope  (GMRT) array.
The GMRT has a ``Y''-shaped hybrid configuration of 30 antennas, each 45 m in 
diameter. There are six antennas along each of the three arms (with arm length 
of 
$\sim $14 km). These provide high angular resolution (longest baseline $\sim $25 km). 
The rest of the twelve antennas are located in a random and compact 
$\rm 1\times1$ $\rm km^{2}$ arrangement near the centre and is sensitive to 
large scale diffuse emission (shortest baseline $\sim $100 m). The observations 
near 1280 MHz were made using the full 16 MHz bandwidth. 
Details of  the GMRT antennae and their  configurations can be
found in  Swarup et al.  (1991).  The source,  3C147, was used  as the
primary  flux calibrator,  while the  source  0432+416 was  used as  a
secondary calibrator  to derive the  phase and amplitude gains  of the
antennae.  The data  analysis was done using AIPS.  The corrupted
data (dead antennae, interference,  spikes, etc.) were identified using
tasks IBLED, UVPLT and VPLOT. The flagging of the bad data was carried
out using  UVFLG and  TVFLG. The image of the field was formed  by Fourier
inversion and cleaning algorithm (IMAGR).

\section {Other Available Data Sets}

\subsection {Near-Infrared Data from $2MASS$}

To complement our optical and NIR imaging data we have used the
data for point  sources around Sh2-294 region from  the Two Micron All
Sky Survey\footnote{This publication makes use of data products from the Two
Micron All Sky Survey, which is a joint project of the University
of    Massachusetts    and    the    Infrared    Processing    and    Analysis
Center/California Institute of Technology, funded by the NASA and the NSF.}
(2MASS)  Point Source Catalog (Cutri et al.  2003) in the J
(1.25 $\mu$m), H (1.65 $\mu$m) and K$_s$-band (2.17 $\mu$m). The 2MASS 
data have positional accuracy of better than $2^{\prime\prime}$.
Our  source selection  was based  on the  `read-flag' which  gives the
uncertainties in the  magnitudes. 
To discuss the radial density profile of the cluster (see \S 4.1) 
associated with Sh2-294 region, we used 2MASS sources with `read flag' values 
of 1-6 from our sample. However to further improve the analysis, only sources  
with `read-flag' values of  1-3  have been used to study the 2MASS 
color-color  (CC)  and color-magnitude (CM) diagrams (\S 4.6). 
It is also to be noted that we used 2MASS images for astrometric
purposes of our optical images with matching tolerance of $1^{\prime\prime}$.0.

\subsection {Mid-infrared data from $MSX$}

The {\it Midcourse Space Experiment}\footnote{This research made use of 
data products from the Midcourse Space Experiment. Processing of the 
data was funded by the Ballistic Missile Defense Organization with 
additional support from NASA Office of Space Science. This research 
also made use of the NASA/ IPAC Infrared Science Archive, which is 
operated by the Jet Propulsion Laboratory, Caltech, under contract 
with the NASA.} 
(MSX) surveyed the entire Galactic plane within $|$$b$$|$ $\le$ 5$^\circ$
in four mid-infrared (MIR) bands
centered at 8.28 $\mu$m (A), 12.13 $\mu$m (C), 14.65 $\mu$m (D) 
and 21.34 $\mu$m (E) at a spatial resolution of  18$^{\prime\prime}$.3 
(Price et al. 2001) and  a global absolute
astrometric accuracy of  about 1$^{\prime\prime}$.9  (Egan et al. 2003).
The MIR data are useful to study the large scale distribution of warm
($\sim$ 1000 K) dust
and  emission from
Polycyclic Aromatic  Hydrocarbons (PAHs).  The MSX A (50\% peak
response range: 6.8 - 10.8 $\mu$m) and C (11.1 - 13.2 $\mu$m) bands
include several PAH features at 6.2, 7.7, 8.7, 11.3 and  12.7 $\mu$m. 
Point sources have  been selected  from MSX  Point Source  Catalog  
Version 2.3 (Egan et  al. 2003) and 
MSX A-band image is used to see  PAH emission around Sh2-294 region.  
The  flux densities  of the MSX point  source associated
with Sh2-294 region are given in Table 2 and used for constructing the
spectral energy distribution (SED).

\subsection {Mid- and far-infrared data from $IRAS$-HIRES}

The data from the {\it Infrared Astronomical Satellite} ($IRAS$) survey in
the four bands (12, 25, 60 and  100 $\mu$m ) for the region around 
Sh2-294 were  HIRES-processed (Aumann  et al. 1990)  at 
the Infrared Processing and Analysis Center (IPAC), Caltech, 
to obtain high angular resolution maps. Apart from determining the flux densities, 
the $IRAS$-HIRES maps have been used to generate the  
dust color temperature and  optical depth maps of interstellar dust. 
The integrated flux densities of the $IRAS$ PSC associated with Sh2-294 region
are given in Table 2.

\section {Infrared and Optical Properties of Stellar Sources}

The H$\alpha$+continuum image and a false $JHK_{\rm s}$ color composite 
infrared image ($J$, blue;  $H$, green; and $K_{\rm s}$, red ) of Sh2-294 
region are shown in the {\it left} and {\it right}  panels of  Fig. 1, respectively.   
A close-up  view of $JHK_{\rm s}$  image reveals  that  there are  two  significant 
amounts  of stellar  density enhancements in  the region.  The eastern  region is
situated at  the border  of H$\alpha$ emission,  hence the  region of
interest of this study.

\subsection{Radial Profile and Stellar Surface  Number Density}

Sh2-294 is listed as a partially embedded infrared cluster 
in the catalog compiled by Bica, Dutra, \& Barbuy (2003).  To  determine  
the boundaries  of  the  cluster,  we
performed direct  star counting, assuming spherical  symmetry. We have
selected  sources which  are  detected  in the  2MASS $K_{\rm s}$ band.  To
estimate   the    cluster   radius    we   selected   a    region   of
$300^{\prime\prime}$ centered on IRAS 07141-0920 source. The center of
the cluster  was estimated  by convolving a  Gaussian kernel  with the
stellar distribution  and taking the  point of maximum density  as the
center.  This  was performed  for  both the  axes  to  get the  central
coordinates   of   the  cluster.   The   center   was   found  to   be at
$\alpha_{2000}=07^{h}16^{m}33^{s}$,
$\delta_{2000}=-09^{\circ}25^{\prime}35^{\prime\prime}$.  To determine
the radial surface density, the Sh2-294 region was divided into a number
of  concentric rings  from the  cluster center.  The  projected radial
stellar density in each concentric circle was obtained by dividing the
number  of stars  in each  annulus  by its  area. The  error bars  are
derived  assuming that  the number  of stars  in each  annulus follow
Poisson  statistics. To account  for the  contribution from  the field
stars,        we         selected        a        control        field
($\alpha_{2000}=07^{h}17^{m}33^{s}$,
$\delta_{2000}=-09^{\circ}25^{\prime}24^{\prime\prime}$)    which   is
$\sim15^{\prime}$ to the  east of IRAS 07141-0920 source. Figure 2
shows the radial density profile  of the cluster (open circles) 
 and the
horizontal  dashed line shows  field star  density estimated  from the
control   field.   A   clear  gradient   in  surface   number  density
distribution of Sh2-294 region confirms the existence of clustering around
the  center of  the region.   Several  studies have  shown that  
the radial density profiles of young embedded  clusters  
can be fitted  by  both  the  inverse  radius
($r^{-1}$)  model and  the  King's model  (e.g. Horner  et al.  1997;
Teixeira et al.  2004; Baba et al. 2004). Neglecting the tidal radius,
the King's profile can be written as 

\begin{equation}
f(r) = a + \frac{f_0}{1+(r/r_c)^2}
\end {equation}

where $r_c$ is the core radius (the radius at which the surface density
falls to half of the central density $f_0$) and $a$ is the constant for
background offset due to field stars. 

                             For  the Sh2-294 cluster,  we  found that
the King's model has a better fit (reduced $\chi^2$ =  0.96) as compared to
`$r^{-1}$' model (reduced $\chi^2$ = 1.04).  Within errors, the King's
profile (dotted curve in Fig. 2) merges with the  background field 
at $\sim2.3^{\prime}$, which
corresponds to  $\sim3.2$ pc at a distance of 4.8 kpc
(see \S 4.5). We take this as the cluster radius.
The background level as estimated from the control
field  is  $\sim$  4.7  stars  arcmin$^{-2}$ which  is  in  reasonable
agreement with  the value of  $\sim$ 3.8 stars  arcmin$^{-2}$ obtained
from King's  profile.

The contour map  of the stellar surface number  density (SSND) around
Sh2-294  region was made using kernel  method (Gomez et al. 1993;
 Silverman 1986) and is  shown in Fig. 3.  The stellar density
distribution     is      smoothened     by     12$^{\prime\prime}$$\times$
12$^{\prime\prime}$ sampling box. At the distance of 4.8 kpc, the unit
cell corresponds  to $\sim$ 0.28  pc $\times$ 0.28  pc. The contour
levels are drawn from 3 times of the  background level with  
an  increment of 2 stars/pc$^2$,
where the background level is taken as 4 stars/arcmin$^2$ (2 stars/pc$^2$).
The peak of the projected stellar 
density distribution was found  to   be  $\alpha_{2000}=07^{h}16^{m}33^{s}$,
$\delta_{2000}=-09^{\circ}25^{\prime}29^{\prime\prime}$,    which   is
close (6$^{\prime\prime}$) to the position  found by convolving  a Gaussian
kernel as  discussed above. The King's profile  fitting 
yields a central  density of 44 $\pm$ 13  stars/arcmin$^2$ ($\sim$ 22
stars/pc$^{2}$), correlates well  with $\sim$ 22 stars/pc$^2$ of the
peak of SSND. The clustering in  the central region is more obvious in
the surface density  map. However, there is evidence that 
Sh2-294 possesses  distinct substructures.  The substructures in  the east
and northwest directions indicate the density enhancement, which 
can also be  seen as a second peak  
in radial density distribution
around  $\sim  1.2^{\prime}$  from   the  central  peak (see Fig. 2).  
The  eastern substructure is roughly spherical 
in shape and more prominent with peak approximately at
$\alpha_{2000}=07^{h}16^{m}38^{s}$                                  and
$\delta_{2000}=-09^{\circ}25^{\prime}40^{\prime\prime}$, 
having central density of 13  stars/pc$^{2}$.  This 
substructure is at a distance of $\sim$ 1.23$^{\prime}$ from the cluster 
center of Sh2-294
region  and is   also  located  beyond  the  faintest  nebulosity
associated  with the  central  cluster  seen in  the  2MASS $K_{\rm s}$  band
image.  Therefore, it  might be  an other independent  cluster  but still
embedded in the  molecular cloud complex harboring Sh2-294. Hereafter, we
refer to this region as  `A'  for our  further discussion  in the
paper.  In Figure  1 ($right$) the `region A' is  represented by a white
circle in $JHK_s$ color composite image.

\subsection{Probable members of the cluster}

To study  the various  parameters of the  cluster, it is  necessary to
remove  field star  contamination from  the  sample of  stars in  the
cluster  region.   Membership  determination  is   also  crucial  for
assessing  the presence  of PMS  stars, because  both PMS stars  and dwarf
foreground  stars occupy similar  positions above  the zero-age
main sequence (ZAMS) in the color-magnitude (CM) diagram. 
In the absence of proper motion, we
used statistical criterion to  estimate the number of probable  field 
 stars in  the cluster  region by selecting  a control region  of same
 area from our CCD image (control field is used to estimate the field star 
contamination). Figures  $4a$ and $4b$ shows $V, (V-I)$ CM diagrams
 for  the cluster and control  field regions, respectively. The
 contamination due to background field population is  clearly visible in
 the cluster region CM diagram.  To  remove contamination of field stars from
 the   cluster  region   sample,  we   
statistically   subtracted  the
contribution of field stars from  the CM diagram of the cluster region
following the procedure given by Mateo \& Hodge (1986). For  a randomly selected star in  the $V, (V-I)$
CM diagram of the field region,  the nearest star in the cluster's $V,(V-I)$
CM diagram  within $V\pm 0.25$  and  $(V-I)\pm 0.125$  of  the  field star  was
removed.  The statistically  cleaned $V,  (V-I)$ CM diagram  of  the cluster
region is shown in Fig. $4c$  which clearly shows the presence of PMS
stars in  the cluster. 
The presence of a few field stars in
the statistically cleaned CM diagram can still be noticed. 
These probable  field stars
are identified by using the CC diagram and marked as triangles
in Fig. 4$c$ as well as in Fig. 6 (see \S 4.4).

\subsection{Spectral type of ionizing source}

The  brightest  member found  in  our  statistically  cleaned CM diagram  is
situated    at     the    center    of     optical    nebula    having
$\alpha_{2000}=07^{h}16^{m}33^{s}$,
$\delta_{2000}=-09^{\circ}25^{\prime}22^{\prime\prime}$.
The  photometric spectral type  of this
star is of B0.5, with $V$ = 13.99 and $B-V$ = 1.03, which is the source
of ionization in Sh2-294 (Moffat, Jackson, \& Fitzgerald 1979). This is in 
agreement with our measurements, $V$ = 14.04, $B-V$ = 1.04 and $M_V$ = -3.56 
at the assumed distance 4.8 kpc (see \S 4.5),
correspond to B0.5 V spectral type (Schmidth-Kaler 1982; Panagia 1973).
The second brightest member of the cluster has $V$ = 15.41
and $M_V$ = -2.20, which is consistent with spectral type between B2-B3 V
(Schmidth-Kaler 1982; Panagia 1973). As stars of 
spectral type B2 or earlier are known to produce a significant proportion
of their radiation as photons with energy high enough to form HII 
region, therefore the star situated at the center is most probably the 
ionizing source of Sh2-294 HII region.

              To determine the spectral type of this ionizing
source, we  extracted 1D-spectrum (see \S 2.2), with  aperture  centered  on 
the star.   Figure  5  shows  the  spectrum  of the  star  in  the  range
3950-4750 \AA. Besides hydrogen Balmer lines (H$_\gamma$,  H$_\delta$,
H$_\epsilon$, etc),
the spectrum displays HeII line at $\lambda$ $\sim$ 4200 \AA, and 
weak HeII line at 4686 \AA, along with OII/CIII blend at 
$\approx$ 4650 \AA, 
indicates that the spectral type of the star is earlier than B1. 
Specifically HeII line at $\lambda$ 4200 \AA~ is seen in stars
of spectral type B0.5-B0.7 or earlier.
On the other hand, the weakness of HeII line at $\lambda$ 4541 \AA~
relative to HeI line at 4471 \AA~indicates an spectral type latter than O9,
while HeII $\lambda$4686 \AA~  $\la$ HeI  $\lambda$4713 \AA~  favors the 
classification of $\sim$ B0.  Comparing the
composite spectrum  with that in  the atlas of  low-resolution stellar
spectra of  Jacoby et  al.  (1984)  and with the  atlas of  Walborn \&
Fitzpatrick (1990), we propose that the exciting source  for Sh2-294
has  an equivalent  spectral  type of  B0 $\pm$ 0.5 V.
 
\subsection{Extinction}

The extinction in star cluster  arises due to two distinct sources; (i)
the general ISM in the foreground of the cluster
($E(B-V)_{min})$,  and (ii)  the localized  cloud associated  with the
cluster  ($\bigtriangleup  E(B-V)= E(B-V)_\star-E(B-V)_{min}$),  where
$E(B-V)_\star$ is  the reddening  of the star  embedded in  the parent
cloud.   The former  component is  characterized by  the ratio  of the
total-to-selective extinction $R_V$ (= $A_V/E(B-V)$)= 3.1
(Wegner  1993; He et  al. 1995; Winkler 1997),  whereas for
the intracluster  regions of young clusters embedded  in dust and gas
cloud the  value of  $R_V$ varies significantly  (Tapia et  al. 1991;
Pandey et  al. 2003).   However, HII regions  associated with  a large
amount of  gas and  dust often show larger values  of $R_V$  (Forte
1978; The \& Groot 1983; Chini \& Wargau 1990; Pandey et al. 2000).

In   the  absence  of   spectroscopic  observation,   the  interstellar
extinction $E(B-V)$ towards  the cluster region can be  estimated using the
$(U-B)/(B-V)$  CC  diagram. The  CC diagram of the cluster region is 
presented in Fig. 6, which contains data points for all the stars having
 V $\le$ 18 mag.
The figure clearly shows  
a large  amount of  contamination due  to foreground  stars. The 
probable field stars in the statistically cleaned CM diagram
follow the sequence of foreground  stars and are marked as triangles
 in Fig. 6. 
The probable cluster members manifest a more reddened
 sequence than the foreground sequence. To match the observations, 
the ZAMS  by Schmidt-Kaler (1982) is shifted along
the  reddening   vector  having  a  normal  slope  of
$\frac{E(U-B)}{E(B-V)} = 0.72$.  The  CC   diagram  indicates  that  the
foreground population  is reddened  by $E(B-V) = 0.2$ mag.  The minimum
reddening for the  cluster region  is estimated to be 
$E(B-V)_{min}$ = 1.35 mag.

\subsubsection {Ratio of Total-to-Selective Extinction}

The  variable  amount  of  absorbing  grains in  the  line  of  sight
determines the  total extinction towards  each star. In  practice, the
color excess for a star is obtained by comparing its intrinsic color
with the observed  color, and the total visual  extinction is derived
by using the shape of  dust-reddening law. The direct method, however,
cannot be easily applied because  the intrinsic colors of most of the
stars are usually unknown, so that an indirect way of statistical approach
can  be adopted  on  the basis  of  number of  stars  detected in  the
investigated region. It has been shown by Chini and Wargau (1990) that
two color diagrams (TCDs) of  the form of $(V-\lambda)$ vs.  $(B-V)$,
where $\lambda$  is one of the  wavelengths of the  broad-band filters
$(R, I, J, H,  K)$,  provide  an effective  method  of separating  the
abnormal  extinction  arising within  such  clusters  from the  normal
extinction produced by general ISM.  Chini \& Wargau (1990) and Pandey
et al. (2000)  used TCDs to study the anomalous  extinction law in the
clusters M 16 and NGC  3603, respectively.

                                     To construct TCDs, we used
the probable  cluster members obtained from  statistically cleaned CM diagram
(cf  section  4.2).   The  stars  showing infrared  excess  are  not
included in the analysis. 
Figure 7 shows the 
$(V-\lambda)/(B-V)$ TCDs
for all the probable members of the cluster region.  The slopes of the
distribution are obtained by using the weighted least-square fit
to the data points. To derive the
value of $R_{cluster}$ (ratio of total-to-selective extinction in the
cluster  region) we use  the approximate  relation (cf.  Neckel \& Chini
1981),

\begin{equation}
R_{cluster} = \frac{m_{cluster}}{m_{normal}} \times R_{normal}
\end{equation}

Here $m_{cluster}$ and $m_{normal}$ are the slopes of the distribution
followed by cluster  stars and normal MS stars (taken from the stellar
models  by  Bertelli  et  al.   1994) respectively. 
The slopes  are  given  in  Table  3.  The
$R_{normal}$  represents   to  the  value   of  3.1. The
weighted  mean   value  of  $R_{cluster}$ comes   out  to  be  3.8 $\pm$ 0.1, 
indicating an anomalous reddening
law for Sh2-294 region. Here we would like to mention that the value of
$R_{cluster}$ obtained by using equation (2) should be considered as an approximate
estimation of $R$ in the cluster region.

To further test the reddening and $R_V$ value, we used the observed 
properties of ionizing source 
and assumed that the spectral type of the star is B0 V.
The observed $V$ magnitude of the ionizing  star  is 14.04 with  $B-V$ = 1.04. 
Comparing with intrinsic colors of Johnson et  al. (1966), the reddening $E(B-V)$ to
the  star  corresponds to  1.35  mag.  The $R_V$ value to  the star  
can be  calculated using  the empirical
relationship established  by Whittet (1976)  between $R_V$,
the color excesses $E(B-V)$ and $E(V-K)$,

\begin{equation}
R_V= 1.10\times E(V-K)/E(B-V),
\end{equation}

To derive $R_V$  for the star, the $JHK_{\rm s}$ magnitudes  of the star are
transformed from 2MASS system  to Koornneef system using the relation
given  by Carpenter (2001)  and intrinsic $V-K$  color  obtained from
Koornneef (1983). The $R_V$ value thus obtained is $\sim$ 3.9. Hence,  
the reddening and ratio of total-to-selective extinction estimated as above
agree quite well with the value found from TCDs.

\subsection{Distance and age of Sh2-294}

The ages of young clusters are typically derived from post-main-sequence 
evolutionary tracks for the earliest members if significant evolution has 
occurred or by fitting the low-mass contracting population with
theoretical PMS isochrones. The fitting of theoretical zero-age main sequence 
(ZAMS) to the observed sequence yields distance. The distance of the cluster 
associated to Sh2-294 is derived using statistically cleaned CM diagrams as  
shown in  Fig. 8. Using the foreground reddening $E(B-V)$ = 1.35 mag towards the cluster 
region and $E(V-R)/E(B-V)$ = 0.60, we visually fitted the 
ZAMS (solid curve) given by Schmidt-Kaler (1982) and Walker (1985) to the 
V, B-V and V, V-R  CM diagrams respectively. The visual fit yields a 
distance modulus of $(m-M)_V$  = 17.6 $\pm$ 0.10, which corresponds to a
distance of 4.8 $\pm$ 0.2 kpc. Our distance estimation is in agreement with 
the most widely used distance of 4.6 kpc for this source.  Theoretical 
  isochrone for 4 Myr ($Z$=0.02) by Bertelli et al. (1994) is also shown  
in Fig. 8. Since massive stars do not show significant evolution,
a maximum age of 4 Myr can be expected for the cluster.
In Fig. 8 we  have also overplotted PMS isochrones from 
Siess, Dufour, \& Forestini (2000) for 1 and 5 Myr. The open circles with dots 
inside represent the optical counterpart of NIR excess stars (see \S 4.6.1), 
which are the probable PMS sources of the cluster. The positions of these 
sources in the CM diagram indicate an age spread of 1-5 Myr.

\subsection {Nature of sources associated to Sh2-294}

In $JHK_{\rm s}$ color composite image (see Fig. 1 $right$), one could see
that the `region A' is  significantly redder  than other  parts  of the
region.  The sources  in this region have no  optical counterpart 
implying that they
are  still deeply  embedded  in  the molecular  cloud  of high  visual
extinction.

\subsubsection {NIR Color-Color Diagram}

In  order to  examine the  nature of  the stellar  population,  we use
$(H-K)$ vs. $(J-H)$ CC diagrams  of the Sh2-294 star-forming  region and the control
field. In Fig. 9 we have plotted  141 and 79 sources for the cluster region
 and the control region, respectively, having  2MASS `read-flag' values of 1-3
(good quality photometric magnitudes)  in all the three $JHK_{\rm s}$ bands.
To plot  the stars in the  CC diagrams, we  transformed the magnitudes
and colors  to the CIT system.   The solid thin  curve represents the
locus  of MS  and thick-dashed  curve is  for giant  stars  taken from
Bessell \&  Brett (1988). The  dotted line represents the classical
 T Tauri (CTT)  locus as  determined by  Meyer, Calvet, \& Hillenbrand (1997).   
The thick
continuous  solid curve  represents the  Herbig Ae/Be  locus  (Lada \&
Adams 1992).  Presuming universal nature  of the extinction law in NIR
region  (Mathis 1990),  we  plotted reddening  vectors  of the  normal
reddening law as  parallel dashed lines.  We assumed  that $A_J/A_V$ =
0.265, $A_H/A_V$ = 0.135, $A_K/A_V$  = 0.090 for CIT system (Cohen et
al. 1981).

The arrow  in Fig. $9a$ points to  the position corresponding  to the
central brightest star in the cluster. We calculated the spectral type
of the star by shifting it along the reddening vector to the intrinsic
main sequence  locus.  We found the  spectral type of  the brightest
star  in the  cluster to  be $\sim$  B0.
In  Figure $9a$ we have shown
the sources belonging to `region  A' in squares; these are the sources in a
circular area of radius  $20^{\prime\prime}$  around  the  second  peak
(eastern substructure) seen in the
SSND diagram (see Fig. 3). Figure $9a$ shows that the average visual 
extinction ($A_V$ $\sim$ 7.4 mag) for the  sources of  `region A'  is  
higher than  the average  $A_V$ ($\sim$ 3.3 mag) of  the sources  
associated to  other parts of Sh2-294 region. We have calculated the extinction
by de-reddening the stars of ``F'' region (see below) in the CC diagram
 to the M0-M6 part of main-sequence locus. 
The high $A_V$ indicates that the  stars in `region A' are more deeply 
embedded thus suffers a higher extinction.

We have classified the CC  diagram into three regions (e.g Sugitani
et al.  2002; Ojha et  al. 2004a,b) to  study the nature of  sources. The
``F'' sources  are located within the  reddening band of  MS and giant
stars. These stars are generally  considered to be either field stars,
Class III  objects or  Class II objects  with small NIR  excess. ``T''
sources  are located  redward  of region  ``F''  but  blueward of  the
reddening line  projected from  the red  end of the CTT locus of
Meyer, Calvet, \& Hillenbrand (1997).   These sources  are considered  to be  mostly T
Tauri stars  (Class II objects)  with large NIR excess.  ``P'' sources
are those located  in the region redward of region  ``T'' and are most
likely  Class I  objects (protostellar  objects).  There  may  be an
overlap in  NIR colors  of upper-end band  of Herbig Ae/Be  stars and
lower-end band  of T Tauri stars  in the ``T''  region (Hillenbrand et
al. 1992).

Majority  of  sources in  our  sample  are  distributed in  the  ``F''
region.  A  total of 36 sources show NIR excess (i.e. sources right to 
the ``F'' region), out of which 18 sources are 
populating in the ``T''  and the  ``P'' regions, characteristic of
young stellar objects (YSOs) of Class II and I. However, this  is the
lower limit as several cluster members detected in the $K_{\rm s}$ band,
may not be detected in the  other two shorter NIR wavelength bands.
The fraction of NIR excess stars in a cluster is an age indicator because 
the disks/envelopes become optically thin with the age. For young 
embedded clusters with age $\leq 1\times 10^6$ yr the fraction is $\sim$ 50\%
(Lada et al. 2000; Haisch, Lada, \& Lada 2000) and the value is $\sim$ 40\%,
as in case of the YSOs
in the Taurus dark clouds, which have an estimated age of about
$\sim 1-2 \times 10^6$ yr (Kenyon \& Hartmann 1995). The fraction decreases 
to $\sim$ 20\% for
the clusters with age $\sim 2-3 \times 10^6$ yr (Haisch et al. 2001; 
Teixeira et al. 2004). 
We have detected 6 point sources in `region A' in all the three bands 
within the 2MASS detection 
limits. One of the sources located left of the reddening vector of giant 
stars (leftmost reddening vector in Fig. $9a$) is probably a field star 
of the region whereas two sources show NIR excess, characteristic of YSOs of 
Class II. The fraction of the NIR excess stars (2/5,  40\%) in  
`region A' thus suggests an upper age limit of $\sim 1 - 2\times10^6$ yr 
of the  cluster. We also found three 2MASS sources detected only in 
$K_{\rm s}$ band within `region A' (see \S 6.1), which could be YSOs in their 
earliest stages still embedded in the cloud.
The non-detection of these sources
in $J$ and $H$ bands limits our analysis to examine the exact nature of each
source.
If we presume that these sources are YSOs having circumstellar disks, 
then the fraction of NIR excess stars (5/8, 62\%) would suggest an 
age of $\leq 1 \times 10^6$ yr to `region A'.
However, to further improve the statistics, observations 
in $L$ and $M$-band are needed.   
In the CC diagram of control field, most of the sources are spread over
lower  part of  ``F'' region;  they  are mainly  foreground stars.  
In comparison the  distribution of  sources in  the cluster  region shows
a mixture  of field  stars  and reddened  sources  probably having  warm
circumstellar dust, characteristic of young PMS objects (Lada \& Adams
1992). In order  to examine the statistical properties  of the cluster
sources, the  histograms of  $H-K$ colors are  shown in Fig.  10.  The
color distribution of sources belonging  to control field is shown in
shaded area, whereas, for the cluster region the distribution is shown
in  thick lines. The histogram shows two peaks in the range 
0.05  $\leq H-K \leq$ 0.2 and 0.35  $\leq H-K\leq$ 0.5 for the cluster region.  
The peak of $H-K$  color distribution  of control region is  ahead of  
cluster region  in the range 0.05  $\leq H-K \leq$
0.2. Hence,  probably the first  peak of  $H-K$ color  distribution of
cluster region may  be due to foreground population  of the region. On
the other  hand, a large number  of sources were found  in the cluster
region  with  $H-K  \geq  0.4$.  The existence  of these reddened  sources
indicates a population of young stars still embedded  in significant
quantities of intervening gas and dust.

\subsubsection {NIR Color-Magnitude Diagram}

Further information about  the nature of the sources  can be extracted
from the $(J-H)$ vs. $J$ and $(H-K)$ vs. $K$ CM diagrams.  In $(H-K)$ vs. $K$ CM diagram (Fig. 11) we have
plotted  all  the 141  sources  detected  in  the $JHK_{\rm s}$  bands  plus
additional 11  sources which are detected  in the $H$  and $K_{\rm s}$ bands
only.  The vertical solid lines  represent ZAMS curves (for a distance of 
4.8 kpc) reddened by A$_V$ =  0,  10 and  20  mag,  respectively. 
The parallel slanting lines
represent the reddening vectors to the corresponding spectral type. An
apparent MS track is noticeable with $A_V \sim$ 1.5 mag in this diagram,
however a comparison of it with a similar diagram for the stars in the
control  field shows that  it is  a false  sequence caused  by field
stars  in the foreground  of Sh2-294 region.
The  foreground sources in cluster region are of $H-K < $ 0.2.
YSOs (Class II and I) found from  the CC diagram (Fig $9a$) are
shown as star symbols and filled triangles,  respectively.  
The probable Herbig  Ae/Be sources are shown as crosses and
the sources  belonging to `region A' are marked in squares. 
Sources detected only  in $H$ and  $K_{\rm s}$ bands
are represented by  large filled circles which may also be YSOs.
For the  sources detected in
$H$ and  $K_{\rm s}$ bands, we have taken the $K_{\rm s}$ magnitude as
K$_{CIT}$ due  to   absence  of  appropriate   transformation 
equation  between
K$_{2MASS}$ to  K$_{CIT}$ system but the  colors are transformed to  the CIT
system.  Out of 11 sources detected  in $H$ and $K_{\rm s}$ bands, only 6 of
them show $H-K \geq 1.0$, hence they are probably young stars in their
earliest evolutionary phases.

The mass of the  probable YSO candidates can  be estimated by
comparing their  locations on the  CM diagram with evolutionary models  of PMS
stars.  We  used $(J-H)$ vs. $J$ CM diagram  instead of the  $(H-K)$ vs. $K$ color-magnitude
diagram to minimize  the effect of the ``excess''  emission in the NIR
to  derive the  ages and  masses of  the YSOs.  Figure  12 represents
intrinsic $(J-H)$ vs. $J$ CM diagram for probable  YSOs (Class II and I). 
The symbols are same as in  Fig. 11. To produce the intrinsic CM diagram, 
the extinction in  front of  YSOs are
derived  by tracing back  along the  reddening vector  to the CTT
locus or its extension in $(H-K)$ vs. $(J-H)$ CC diagram (Fig. 9a).
Here we would like to mention that the present derivation of 
extinction is an approximate approach.
 The thin solid line
represents PMS isochrone of 1$\times 10^6$  yr from 
Siess, Dufour, \& Forestini (2000), at
an assumed distance of 4.8 kpc.  Arrow lines are the reddening vectors
for  sources with  masses of 0.5 and 3.5 M$_\odot$  respectively for
1$\times 10^6$ yr. As can be seen in Fig. 12 the majority of YSOs have 
masses  in the range of 0.5 to 3.5 M$_\odot$ indicating that these may be 
T-Tauri stars.

\section{Description of the HII region}

\subsection {Emission from  Ionized Gas}

Figure 1 shows the H$\alpha$+continuum image of Sh2-294 region. The
nebula looks  loose bipolar in  nature, having an  elongated obscuring
dark-lane  extending  approximately in  the  east-west direction  cutting
across  at  the  center.  The  obscuring  dark  lane  at  the  center
suggests  the  presence  of  a  molecular cloud.   Indeed  Blair, Peters,
\& Vanden Bout (1975)  detected CO (J = 1 $\rightarrow$ 0) emission over a
$6^{\prime}\times6^{\prime}$  area  with   peak  in  CO  intensity  at
$\alpha_{2000}=07^{h}16^{m}32^{s}$,
$\delta_{2000}=-09^{\circ}25^{\prime}23^{\prime\prime}$.   The   lines
were centered at V$_{LSR}=+32$ km $s^{-1}$.  The dark-lane divides the
nebula into  two parts, with bright  nebulosity to the  south and less
bright to the north,  which decreases in surface brightness  away from the
obscured region.  The southern component is brighter and has sharp arc
structure  in the  east  direction giving  an  impression of  possible
ionization  front  (IF).   Whereas  the  northern  component  is  more
extended  and very filamentary in structure, it gives  a stunning  view  of the  dust
content of  the region, proving that  a significant amount  of dust is
concentrated  in the  string-like  filaments across  the region.  This
structured feature  suggests an inhomogeneous  spatial distribution of
the interstellar matter.

Felli \&  Churchwell (1972)  reported the radio  continuum emission
from Sh2-294 region in their low resolution (HPBW = 10$^{\prime}$) 1.4
GHz survey  of Galactic  HII regions. Because  of this  resolution the
object appeared featureless.
 However, the
observation    with    VLA   (Fich    1993)    with   resolution    of
$40^{\prime\prime}$ shows that there  is a  gradient in free-free
emission surface  brightness from southeast  to northwest part  of the
nebula. Figure 13 represents the radio continuum image of Sh2-294 with
GMRT  at 1280  MHz.  The radio  continuum  image has  a resolution  of
33$^{\prime\prime} \times  17^{\prime\prime}$. The map  has an rms noise
of $\sim 0.26$ mJy/beam. The integrated flux  density in the  map is of
$\sim  0.4$ Jy, obtained  by integrating  down to  3$\sigma$  level,
 where $\sigma$ is the rms noise in  the map. The radio morphology shows that
the  thin ionization-bounded  zone  in south-eastern  direction is  an
indication of  IF. 
The decreasing  intensity distribution on the opposite
side  signifies  the region  could  be  density  bounded in  northwest
direction. From  the optical and radio morphologies,  it
appears  to be the  flow of  ionized gas  in northwest  direction, as
expected  in case of  champagne model  (Tenorio-Tagel 1979).
Figure  14 represents the  radio contours of GMRT  at 1280
MHz superimposed on the $K_{\rm s}$  band image of 2MASS in logarithmic scale.  
The bright source
which is the possible source of ionization is located at the geometric 
center of
the radio map and the embedded sources of the `region A' are seen in the
direction of IF,  which could be the probable young PMS sources  of the Sh2-294
region.         The        radio        emission       peaks        at
$\alpha_{2000}=07^{h}16^{m}37^{s}$,
$\delta_{2000}=-09^{\circ}25^{\prime}45^{\prime\prime}$.  At  the peak
of radio emission, we did  not see any significant amount of H$\alpha$
emission, which could be  due to presence of dense ISM/cloud which
obscures the optical emission.

                    Assuming  that  Sh2-294  is  roughly  spherically
symmetric ionization   bounded,  and,  neglecting  absorption  of
ultraviolet radiation by dust inside the HII region, the measured flux
density  of 0.4 Jy,  together  with  assumed  distance, allow  us  to
estimate the  number of  Lyman continuum photons  ($N_{Lyc}$) emitted
per second, and  hence the spectral type of  the exciting star.  Using
the  relation  derived  by  Lequeux  (1980)  and  
assuming an electron
temperature   of  8000 K,  we   estimate
log$N_{Lyc}$ =  47.90, which  corresponds to  a MS  spectral  type 
earlier than  B0 (Panagia 1973)  and close to  B0.5 (Vacca, Garmany, \& Shull 1996;
Schaerer \& de  Koter   1997),  respectively. The differences in the
spectral type is expected due to
uncertainties  in the  effective  temperature and  ionizing fluxes  of
massive stars, for  a given spectral type.  

\subsection{Dust continuum emission}

The strong mid- to far-infrared continuum emission detected from the
local region of the interstellar cloud near `region A' has been 
modelled using a radiative transfer scheme (Mookerjea \& Ghosh 1999).   
The energy from the central embedded source is transported through
the dust and gas (only hydrogen) components of the spherical cloud
and the emergent spectrum is compared with the observed SED (see Table 2). 
The best fitting model is compared with the IRAS and MSX data in Fig. 15
(solid curve).
The far infrared emission has been integrated over a 3$^{\prime}$ (diameter)
region centered on IRAS 07141-0920 from the $IRAS$-HIRES maps.
There is however an excess emission at 12 $\mu$m, which could be 
due to a component of very small grains or polycyclic aromatic hydrocarbons
(Degioia-Eastwood 1992).
The fit is reasonably good. The parameters of this model are : $r^{0}$ 
density 
law, radial optical depth of 5.0$\times$10$^{-2}$ at 100 $\mu$m, central 
dust-free cavity size of 0.012 pc, total luminosity corresponding to a ZAMS 
star of type B1.5. It may be noted that the predicted radio emission at 
1280 MHz is too low compared to that observed locally (170 mJy within
3\arcmin~ diameter). With the aim of reproducing the observed radio emission, 
another model has been constructed. In order to fit the radio data well, 
an entirely different type of the embedded ZAMS star (O6.5) was necessary. 
The resulting emergent spectrum in this case is also presented for 
comparison in Fig. 15 (dashed curve) which is at variance with the observed 
infrared emission. 
Hence, we propose that while the continuum 
emission from the dust component is powered by the embedded source, 
which might be so heavily obscured that it emits mainly in  mid- to 
far-infrared, the ionization of the gas is maintained externally (the 
B0 star discussed in \S 5.1).

\subsection {Photodissociation region}

Figure 16 shows the continuum-subtracted H$_2 (\nu) = 1- 0 S(1)$ 2.12
$\mu$m image smoothed with a 3 $\times$ 3 boxcar. The image has a
field of view of 3.6$^{\prime}$ $\times$ 3.6$^{\prime}$. 
Due to low signal-to-noise ratio, we  can roughly describe
only  some  of  the   characteristics  of  H$_2$  emission.  There  is
significant amount of diffuse H$_2$ emission seen at the center of the
nebula, where the dark-lane is  located, which is correlated well with
the  CO  peak. This kind of structure has been observed in the star-forming 
region S255-2, where the central core of S255-2 is immersed in a 
region of H$_2$ emission with marginal evidence for
limb brightening to the south (Howard, Pipher \& Forrest 1997). 
We also found an  arc-shaped  diffuse H$_2$  emission
feature towards  east, as seen in case of Sh2-219 
(Deharveng  et al.  2006) in the direction  of IF, which could be
 due to the interaction of massive star(s)
with the nearby  molecular cloud.  The diffuse H$_2$  emission can be
excited  by  Lyman  and  Werner  UV  photons  (e.g.,  Chrysostomou  et
al. 1992) from the massive  stars and thus can trace photodissociation
(PDR)  regions;  but for  a  dense  gas  ($\large{n}_{H}$ $\sim$ 10$^6$
cm$^{-3}$), H$_2$ 
emission alone cannot be  used to
identify  PDR  and hence  another PDR  tracer like  Polycyclic Aromatic
Hydrocarbon  (PAH) emission is  necessary to  trace PDR  around nebula
(Giard et al. 1994).    

The far-UV radiation (6 $\leq h\nu  \leq 13.6~eV$) can escape from HII
region and penetrate  the surface of molecular clouds,  leading to the
formation of PDR  in the surface layer of the  clouds. PAHs within the
PDR  are  excited  by  the  UV photons  re-emitting  their  energy  at
MIR wavelengths,  particularly between 6 - 10 $\mu$m. The 
A-band (8 $\mu$m) of MSX includes several  discrete PAH emission features  
(e.g. 6.6, 7.7 and  8.6 $\mu$m) in addition to the contribution from thermal 
continuum  component from hot
dust.  Figure  17 shows H$\alpha$  image superimposed with  MSX A-band
contours. The  8 $\mu$m emission displays partial ring  in eastern side
of Sh2-294 region, with  diffuse emission extending towards  the dark
lane of optical  nebula. The ring of PAH emission  lies beyond the IF,
indicating the  interface between the  ionized and molecular  gas. The
absence  of  8 $\mu$m  emission  in  the  interior  of  HII  region  is
interpreted  as  the  destruction  of  PAH  molecules  by  intense  UV
radiation  of the  ionizing  star.  The peak  of  PAH emission  
matches closely
with diffuse   H$_2$  emission seen in the east.
 While  comparing  the  PAH emission with H$\alpha$ emission, 
it appears that the diffuse  emission of MSX  corresponds to dark optical
  features of H$\alpha$ emission.

\subsection{Dust optical depth and temperature distribution}

The temperature of interstellar dust  is determined by a
balance  between  radiative heating and cooling processes.  Heating occurs  primarily  by
absorption  of stellar  UV  photons, and  cooling  by re-radiation  in
mid- and far-infrared. A dust  color temperature of each pixel  using the flux
ratio at  60 and  100 $\mu$m can  be determined by  assuming a
dust  emissivity law.   We  have  used HIRES  processed  $IRAS$ maps  to
generate maps of the dust color temperature ($T(60/100)$) and optical
depth ($\tau_{25}$  and $\tau_{100}$) around  the Sh2-294  region. The
intensity  maps  at 12,  25, 60  and  100  $\mu$m were  spatially
averaged before computing  $T(60/100)$ and ($\tau_{25}$ ,$\tau_{100}$)
in  a  similar  manner as  described  by  Ghosh  et al. (1993)  for  an
emissivity law of $\epsilon_{\lambda} \propto \lambda^{-1}$.

The HIRES processed $IRAS$ maps at 12, 25, 60 and  100 $\mu$m are shown
in Figure 18.  The infrared emission  in 60 and 100 $\mu$m maps
peaks              at              $\alpha_{2000}=07^{h}16^{m}37^{s}$,
$\delta_{2000}=-09^{\circ}25^{\prime}51^{\prime\prime}$.    
The peak of infrared emission coincides ($\sim 9^{\prime\prime}$)
with the 1280 MHz GMRT radio continuum  peak and lies $\sim 16^{\prime\prime}$
away from the  peak of central density found in SSND distribution
(see \S 4.1) for `region A'.  
The maps  at 60  and  100 $\mu$m trace  the  distribution of
colder  dust. The  peak of  cold dust,  along with  the peak  of radio
continuum suggest that the gas-dust coupling is rather efficient in the
`region A', which is normally the case for  
regions of recent star  formation.  The
dust emission  seems to favor  southeast to  northwest orientation 
 as for the cloud  of Sh2-294 region. The extensions of the 60 and
100 $\mu$m  dust emission maps have diameters  larger than 12  and 25 $\mu$m
maps. This  is due to the fact that  the former has its  main contribution from
the cooler component of dust grains lying in the outer envelope of the
cloud. 
Figure  19 shows the  contour maps  of dust
optical  depth  at 25 \&  100  $\mu$m.  The  $\tau_{100}$  map
represents  low optical  depth at  the  position of  MSX point  source
and/or  `region  A', however two peaks (with a peak value of 
$\tau_{100}$ $\sim$  0.001) can be seen at the
positions along northeast  and southwest of `region A', indicating presence
of high densities in the eastern side of Sh2-294.  A gradient is
seen in the optical  depth map from east to  northwest direction.  The low
optical depth in the northwest direction may be due to the lower dust column
density along the path.
Figure 20  presents a
contour  map   of  $T(60/100)$.   
The  dust color temperature varies from  30 to 40 K in the map.
The $T(60/100)$ and $\tau_{100}$ distributions suggest an anticorrelation,
such as   
the peak position  of $\tau_{100}$ map  corresponds to a
low color temperature of $\sim$  31 K. The relatively high temperature
around `region A' may probably be due to the radiation from young stars of the region.
 The cometary morphology of the
map  could be due  to inhomogeneous  distribution of  dust in  Sh2-294
region.

\section {Star formation scenario around Sh2-294}

\subsection {Spatial distribution of young sources}

Figure   21   shows   the  continuum-subtracted   molecular   hydrogen
2.12 $\mu$m  image of field of view 3.6$^{\prime}$ $\times$ 3.6$^{\prime}$, 
overlaid  by  the  GMRT radio  continuum
contours  (blue color).   The red  contours represent  the  MSX 8 $\mu$m 
emission,  which are generally  attributed to PAHs. 
  The  spatial  distribution  of  Class II,  Class  I,
 probable Herbig  Ae/Be sources and  sources detected only in  $H$ and
 $K_{\rm s}$  bands are  shown in  different  symbols. The  symbols for  the
 sources are  same as in $(H-K)$ vs. $K$  CM diagram (Fig. 11). 
We also find that there are five sources detected only in the
$K_{\rm s}$ band but not in $H$ and $J$ bands.
These sources may  also be
 probable YSO  candidates; we represented them by red filled
 circles  in Fig.  21.  We  found a non-uniform  distribution of  young
 sources in Sh2-294 star-forming  region. The complex seems to be
composed of sources of different evolutionary stages aligned
from northwest to southeast along the dark lane seen  in H$\alpha$ 
image (see Fig.  1). They are  mainly concentrated around the  region where
 diffuse  H$_2$   emission  is  seen.
  Most of the
 Herbig Ae/Be and Class II
  sources   are seen in northwestern direction but in contrast, sources 
 detected in  $H$ and $K_{\rm s}$  bands and only in $K_{\rm s}$ band are seen towards
 southeastern direction. The spatial distribution suggests that the
 southeastern region is less evolved than other parts of Sh2-294 star
forming region.  The   regions  (northeast  and
 southwest) which are most probably ionized by a massive star located
 at the  center of  the nebula, have lack  of Class  II, Class I  and other
 probable YSOs,  which implies that star  formation activity 
has subsided  in these directions due  to intense  UV photons.  
We   identified    a    source   from    MSX   PSC  at
 $\alpha_{2000}=07^{h}16^{m}38^{s}$,
 $\delta_{2000}=-09^{\circ}25^{\prime}40^{\prime\prime}$  
which  coincides with the peak by found  SSND distribution 
for `region A'.  
In Sh2-294 region the young sources in and 
around `region  A' are observed in PDR region along the direction of  IF.  The  
morphology of the PDR and H$_2$  emission in comparison with radio continuum  emission 
gives evidence of interaction  of massive star(s)  with nearby cloud.  The close  
association of  the  partially embedded  cluster (`region A') adjacent to  HII  
region in the direction of IF therefore suggests that the star formation at  the 
border of Sh2-294 may be due to expansion of HII region. It is therefore necessary to 
invoke the classical triggering  scenario of Elmegreen  \& Lada  (1977) that
UV radiation compressed the abundant nearby gas, leading to the star formation.

\subsection {Age of the ionizing source and the HII region}

The spectral type of  ionizing star is of  B0 $\pm$ 0.5 V (see \S 4.3),
which agrees  well  with the  spectral  type  derived  from radio
continuum emission  at  1280  MHz 
and also from NIR CC diagram. Therefore, we assume the
spectral type of the massive star of B0 V for further discussion.  
The probable age ($\sim$ 4$\times10^6$ yr) of ionizing star 
compared to the MS life time of B0 type star ($\sim$  1 $\times 10^7$ yr),
implies that the  HII region is  still under expansion.
The dynamical  age of Sh2-294 can  be estimated by using  the model by
Dyson \&  Williams (1997),  assuming ionized gas  of pure  hydrogen at
constant temperature in an uniform medium of constant density.  If  
the HII region associated to Sh2-294 is excited  by a  B0 V star, 
which  emits $\sim$  78.75 $\times$10$^{46}$ (Log$N_{Lyc}\sim$ 47.90)  
ionizing photons per second and is expanding  into a homogeneous medium  
of density 10$^3$ (10$^2$) cm$^{-3}$,
 will form a Str\" omgren radius 
(r$_s$) of $\sim$  0.17 (0.79) pc.  The initial phase is 
overpressured compared to the neutral  gas and will expand into the medium
of radius r$_i$ at time $t$ as

\begin{equation}
r{_i}= r{_s}(1+{7C_{II}t \over 4r_s})^{4/7} ,
\end{equation}

The observed  diameter of the ionized  gas from GMRT  radio contour is
$\sim$ 4.7$^\prime$, which corresponds to a physical diameter of about 6.6 pc
at a distance of 4.8 kpc.
An assumed sound speed ($C_{II}$)  of 11.4 (4.0) km $s^{-1}$ in the ionized  gas 
of density 10$^3$ cm$^{-3}$ yields dynamical  age of 
 $\sim$  1.5 (4.2) $\times 10^6$ yr,
 whereas in the lower density medium (for example 10$^2$ cm$^{-3}$) the above 
velocity range will yield $\sim$ 0.4 (1.2) $\times 10^6$ yr.
 However the
uncertainty  in  age  estimation  can  not be  ignored  because  of
the assumption of a uniform medium  and  also the  velocity  of expansion
changes  with time.   One would then expect an expanding  IF  to be
preceded by a swept-up shell of cool neutral gas as it erodes into a 
neutral cloud. We have detected a half ring of MSX dust emission towards the
eastern part of HII region which is an indication that neutral gas
surrounds the ionized gas.

\subsection{Triggered star formation}

Triggered  star  formation is  a  sequential  process,  in which  star
formation at  the interface between  HII regions and  molecular clouds
can be triggered by the ionization or shock fronts produced by massive
stars.  The mechanisms mainly responsible for triggered star formation
have been  discussed such as: the ``radiation  driven implosion" (RDI)
and the ``collect-and-collapse" model. In the RDI model, the expanding
ionization front compresses the  pre-existing molecular  clumps, leading to
density  enhancements, which  when it exceeds  the local  critical mass,
collapses  to  form new  stars  (Lefloch  \&  Lazareff 1994;  Gorti  \&
Hollenbach   2002).   
In the ``collect-and-collapse" model, the
neutral material  accumulates between the ionization front  of the HII
region  and the  shock  front  in the  neutral  gas. This  compressed,
shocked   material   may   be   dynamically  unstable   on   a   short
internal-crossing  time scale,  leading to  the formation  of cometary
globules  and bright  rims  (Garcia-segura \&  Franco  1996) or  the
compressed  layer  may   become  gravitationally  unstable  along  its
surface,  on a long  time scale  (Elmegreen \&  Lada 1977;  Zavagno et
al. 2006), leading to the formation of massive fragments forming massive
stars and/or clusters.

The CM diagrams (Fig. 8) indicate that ionizing source may have a maximum age 
of $\sim 4\times10^6$ yr.
It is rather difficult to  
estimate the age of the cluster associated with `region A' as it has no 
optical counterpart and has poor statistics,
however the age to the `region A' can be assigned on the basis of
the age of CTT stars.  The life time  of CTTs ranges from a few 
$\times 10^5$ to few  $\times 10^6$  yr.  The `region A' contains  two CTT  
stars and their positions  in intrinsic $(J-H)$ vs. $J$ CM diagram  
suggest an age of $ < 1\times10^6$ yr. However, this age 
should be considered as an approximate estimation because of uncertainty
in determination of  extinction as well as error in colors.
On the basis of the fraction of NIR excess stars (cf. \S 4.6.1), an age of 
 $\sim$ 1$\times10^6$ yr can also  be assigned to the `region A'.
If we assume that the distance between the ionizing star and
the young stars  in the `region A' is closed to  a projected distance of
$\sim$  1.8 pc (1.3$^{\prime}$) and  that  the ionization  front
expands  at  a  sound  speed  of 4.0 (11.4)  km s$^{-1}$,  it  would  take
approximately 4.6 (1.7) $\times$ $10^5$  yr to reach the surface  of the 
cloud and can induce second generation stars at the border.
The protostellar phase in the clouds  last for $\sim 10^5 $
yr. Therefore, the age difference  between the young sources in 
`region A'  and  central   ionizing  source,   
keeping in mind the uncertainty in assigning the ages to the clusters, seems 
to be consistent with the hypothesis that star formation at the border
may be due to RDI process.
  
However, the possibility of ``collect-and-collapse'' process can not also
be ruled out by considering the following evidence. 
The  radio observation  reveals that Sh2-294  is a roughly spherically
ionized region and is surrounded by  half ring of MSX dust emission in
the mid-IR  (see Fig. 17). The ring  is enhanced in the  middle of the
arc of  dust emission. We  found a MSX  point source at the  center of
dust  peak.  The presence  of dust  ring beyond  the IF indicates
that the neutral  gas surrounds  the HII region.  An embedded  cluster 
(`region A') is associated with  the dust ring  within 2MASS detection limit and
its central density lies close to  the position of MSX point source,  
satisfying most of the predictions proposed by  
the Deharveng et al. (2005) for star formation to occur at the
border due to ``collect-and-collapse'' model.
For example the morphology of Sh2-294 region is almost similar to
the selected HII candidates by  Deharveng et al. (2005), which are likely
to be examples of star formation at the border by this process.
However to look for the distribution of
dense fragments around the Sh2-294, as seen in case of RCW 79 (Zavagno et
al. 2006), one needs millimeter observations.  The fact that the
dynamical age of  HII region is of $\sim$ 1.5$\times10^6$ yr, comparable to
the age  of the stars  associated to the `region A' and the 
presence of a ZAMS B1.5 star further supports the formation of massive star  
following the collapse of thin shell of neutral matter,
as predicted by  ``collect-and-collapse'' model.

With the existing observations and data, it will be  early to establish
or rule out either of the scenarios. To provide a conclusive answer to
the above mentioned star formation scenarios in Sh2-294 we need deep optical 
and  NIR observations, CO, as  well as  millimeter observations. 

\section{Summary}

In this paper we provide the first global view of star formation around
Sh2-294 region. Our multiwavelength data provides a detailed insight into
distribution and nature of the young stellar objects as well as into the
morphology of the  thermal ionized gas and dust emissions. Based on the 
2MASS data, we identify a star cluster at the center of Sh2-294 and a 
distinct group of reddened stars (small cluster) located at  1.8 pc to the 
east of Sh2-294.
The optical and near-infrared color-color and color-magnitude diagrams 
have been constructed to estimate 
the ratio of total-to-selective 
extinction, reddening, distance
and spectral types of young stellar objects. 
The narrow band imaging in H$\alpha$ and radio continuum map based on GMRT 
observation at 1280 MHz have been used to trace the ionized gas, which show
interesting morphological details, including an arc-shaped structure in
southeast direction and creation of lower density zones in the northwest
direction through which the ionized gas can escape. We identified the most
luminous member of the cluster situated at the center of the ionized gas and 
derived its spectral type which is consistent with 
an $\sim$ B0 main sequence star. The number of Lyman continuum photons
from the radio flux suggests that this single star is responsible for
the ionization of the nebula. The PDR at the interface between the ionized
gas and the molecular cloud is traced to the southeast direction by half
ring of PAH feature in MSX and diffuse H$_2$ emission.
The morphological correlation between ionized gas, PDR and H$_2$ emission
leads us to conclude that there is a high degree of interaction of UV 
photons from B0 star with the nearby molecular cloud in north eastern 
direction.  The small cluster is found at the peak of dust ring.
The spatial distribution of the YSOs (Class II sources, sources detected in 
$H$ and $K_{\rm s}$ bands and only in $K_{\rm s}$ band) indicates that 
the region around a small cluster is less evolved and suggests that the
star formation activity observed at the border of Sh2-294 region is
possibly triggered by the expansion of HII region.  

\acknowledgments
We thank the referee for useful comments and suggestions, which  
improved the contents of the paper. The authors would like to thank the staff of HCT at IAO, 
Hanle and its remote control station at CREST, Hosakote, for their assistance 
during observations. The facilities at IAO and CREST are operated by the 
Indian Institute of Astrophysics, Bangalore.
We thank the staff at the GMRT who have made the radio observations 
possible. GMRT is run by the National Centre for Radio Astrophysics of 
the Tata Institute of Fundamental Research (TIFR). MRS would like to thank
the TIFR for the kind hospitality during his visits at the institute, where
a part of the work reported here was carried out. AKP is thankful to the 
National Central University, Taiwan and TIFR, Mumbai, for the financial support 
during his visit to NCU and TIFR respectively.

\clearpage

\begin{table}
\caption{Log of observational data.}
\begin{tabular}{llll}
\\
\tableline
\tableline
Date (UT)   &  Filter     & No of & Exposure (sec)\\
& &\multicolumn{1}{c}{frames} & \multicolumn{1}{c}{per frame} \\
\tableline
2005 Dec 26 &\hspace{2mm} U & \hspace{2mm}    3         &\hspace{4mm} 1200\\
2006 Mar 06 &\hspace{2mm} U & \hspace{2mm}    2         &\hspace{4mm} 300\\
2005 Dec 26 &\hspace{2mm} B & \hspace{2mm}    3         &\hspace{4mm} 600\\
2006 Mar 06 &\hspace{2mm} B & \hspace{2mm}    2         &\hspace{4mm} 240\\
2005 Dec 26 &\hspace{2mm} V & \hspace{2mm}    3         &\hspace{4mm} 600\\
2006 Mar 06 &\hspace{2mm} V & \hspace{2mm}    2         &\hspace{4mm} 180\\
2005 Dec 26 &\hspace{2mm} R & \hspace{2mm}    3         &\hspace{4mm} 300\\
2006 Mar 06 &\hspace{2mm} R & \hspace{2mm}    2         &\hspace{4mm} 120\\
2005 Dec 26 &\hspace{2mm} I & \hspace{2mm}    3         &\hspace{4mm} 300\\
2006 Mar 06 &\hspace{2mm} I & \hspace{2mm}    2         &\hspace{4mm} 80\\
2006 Sep 28 &\hspace{2mm} H$\alpha$ &  \hspace{2mm}    1         &\hspace{4mm} 900\\
2006 Sep 28 &\hspace{2mm} H$\alpha$ {\it cont} & \hspace{2mm}    1         &\hspace{4mm} 900\\
2005 Nov 28 &\hspace{2mm} H$_2$ & \hspace{2mm}  21         &\hspace{4mm} 90\\
2005 Nov 28 &\hspace{2mm} K$_{cont}$ & \hspace{2mm}   21         &\hspace{4mm} 30\\
\tableline
\end{tabular}
\end{table}

\begin{table}
\caption{Flux density details of the $MSX$ and $IRAS$ point sources}
\vspace{0.5cm}
\begin{tabular}{|c|c c c c|c c c c|}
\tableline
\tableline
&  \multicolumn{8}{c|}{Flux Density (Jy) for $\lambda$ ($\micron$)} \\ \tableline
Source & \multicolumn{4}{c|}{$IRAS$-HIRES image$^a$} & \multicolumn{4}{c|}{$MSX$ PSC} \\
       & \multicolumn{4}{c|}{$IRAS$ PSC}              & \multicolumn{4}{c|}{ } \\ \hline
\tableline
       & 100 & 60 & 25 & 12 & 21.3 & 14.7 & 12.1 & 8.3 \\
\tableline
IRAS 07141-0920 & 240.30 & 125.50 & 18.03 & 10.00 & ... & ... & ... & ... \\
                & 262.40 & 77.55  &  7.17 &  2.82 & ... & ... & ... & ...\\
MSX6C G224.1880+01.2407 & ... & ... & ... & ...& 3.40 & 0.86 & 1.90 & 0.88\\
\tableline
\end{tabular}

$^a$ Fluxes obtained by integrating over a circular region of diameter $3\arcmin$ centered 
on the peak. \\

\end{table}

\begin{table}
\caption{Slopes of star distribution in Sh2-294}
\begin{tabular}{clc}
\\
\tableline \tableline \noalign{\smallskip} Color ratio & $m_{cluster}$ & $m_{normal}$ \\
\tableline
\noalign{\smallskip}$\frac{(V-R)}{(B-V)}$&-0.62$\pm$
0.02&-0.55   \\
\noalign{\smallskip}$\frac{(V-I)}{(B-V)}$&-1.30$\pm$ 
0.04&-1.10   \\
\noalign{\smallskip}  $\frac{(V-J)}{(B-V)}$&-2.67$\pm$
0.07&-1.96   \\
\noalign{\smallskip}  $\frac{(V-H)}{(B-V)}$&-3.10$\pm$
0.08&-2.42  \\
\noalign{\smallskip}  $\frac{(V-K)}{(B-V)}$&-3.40$\pm$
0.11&-2.60  \\
\noalign{\smallskip} \tableline
\end{tabular}
\end{table}

\clearpage

\begin{figure}
\epsscale{1}
\plottwo{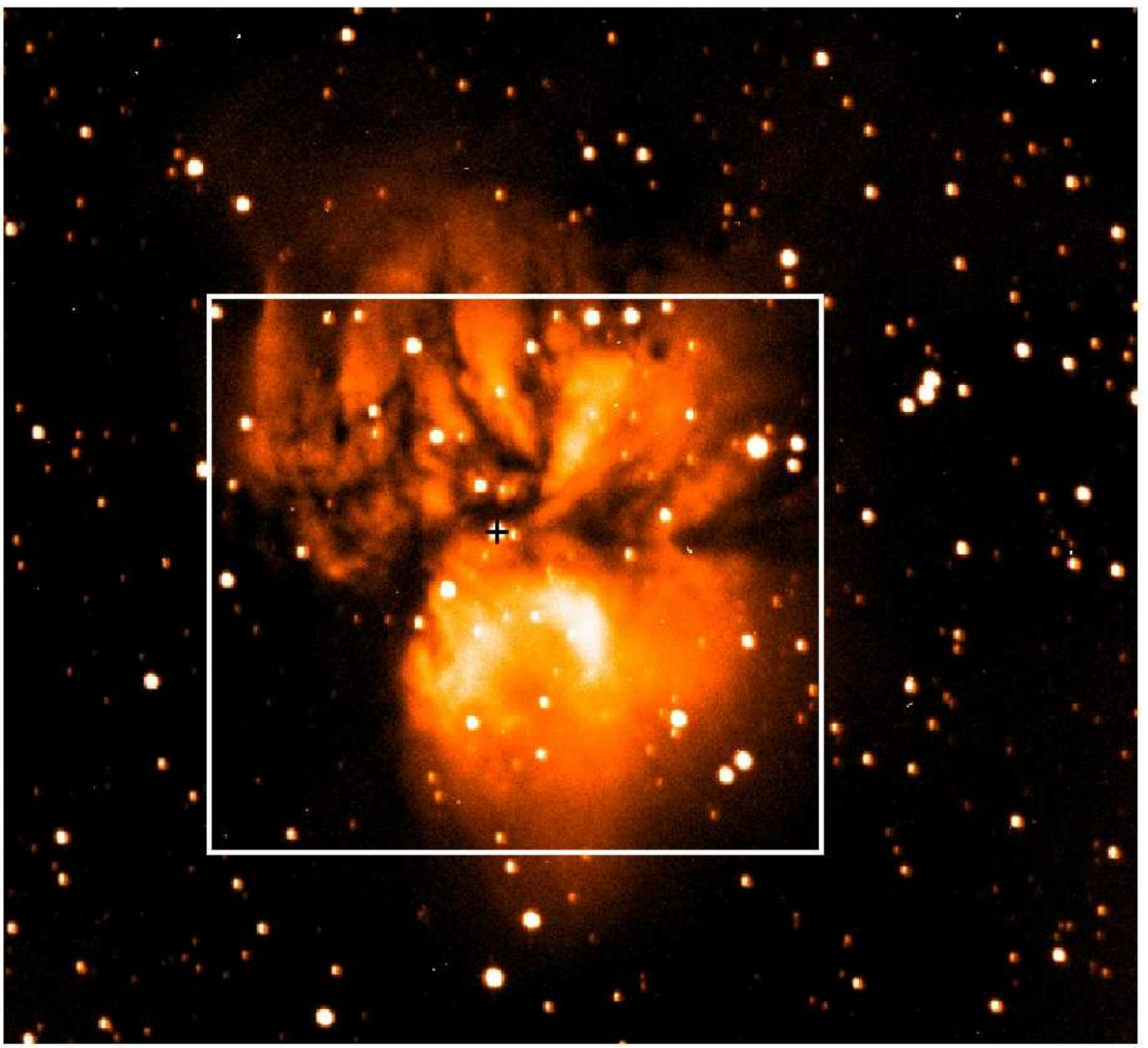}{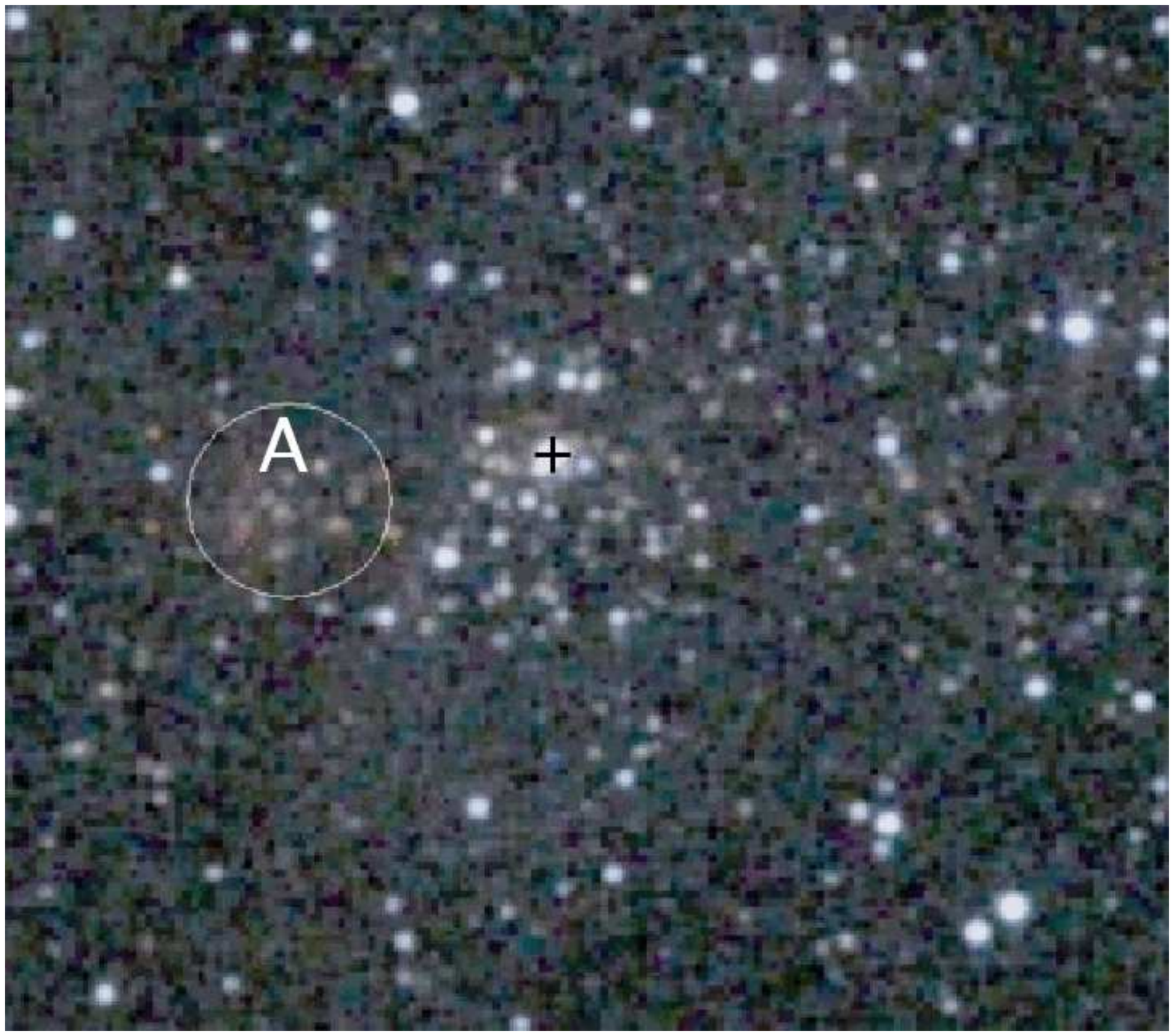}
\caption{$(Left$) H$\alpha$+continuum image ($\sim$ 9.7\arcmin x 8.6\arcmin)
of the ionized gas associated with
Sh2-294 region. The  box in the H$\alpha$ image shows the FOV of
the NIR image shown in
the right panel. ($Right$) Corresponding $JHK_{\rm s}$ color composite image
($\sim$ 5.1\arcmin x 4.7\arcmin) from
2MASS data. North is up and east is to the left. The circle in the NIR image
indicates the young embedded cluster at the border of HII region
(see text for details). The plus sign in both the images marks the position of
possible exciting source of Sh2-294 region (see \S 4.3).
\label{fig1}}
\end{figure}

\begin{figure}
\vspace*{-5.5cm}
\plotone{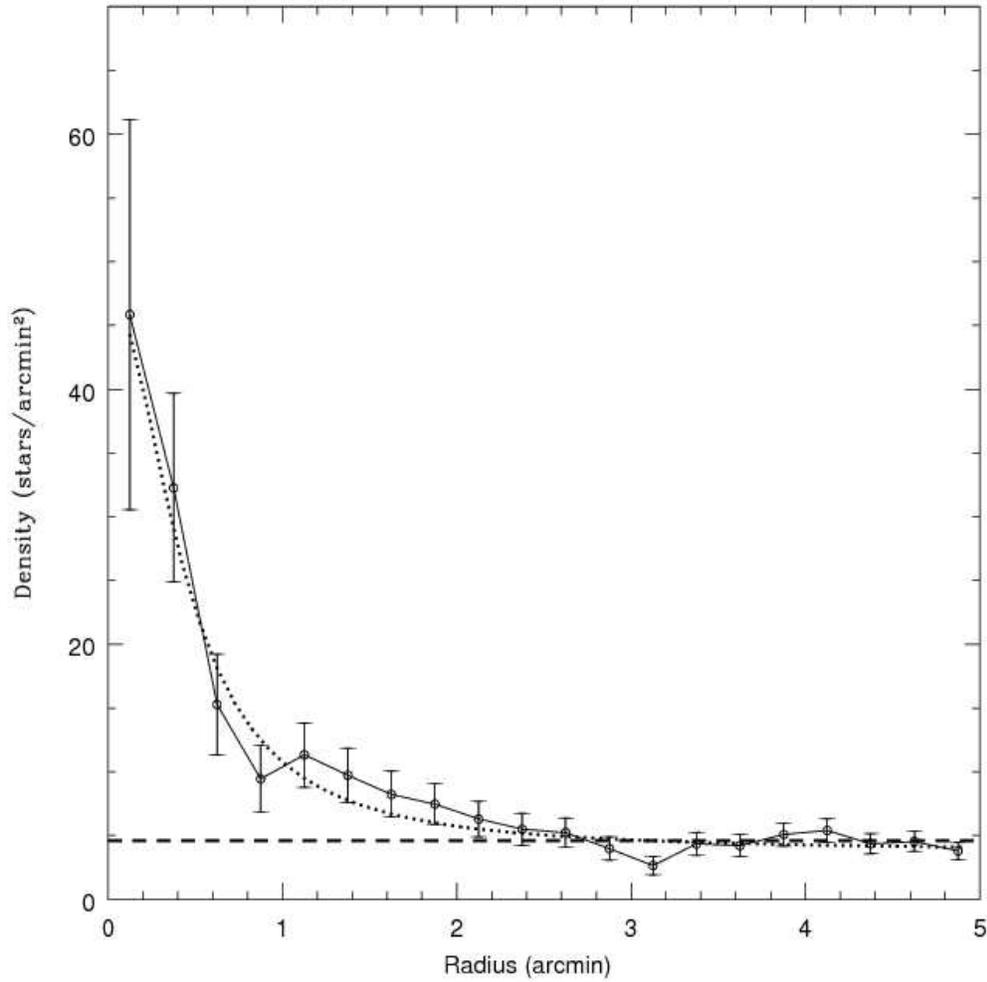}
\vspace*{-1.cm}
\caption{Surface number density profile as a function of radius for
the sources associated with Sh2-294 region. The fitted King's profile
(dotted line) and the radial variation of projected stellar density (solid line)
indicates that the cluster is confined within 2.3$^\prime$ radius. The error
bar represents $\pm \sqrt{N}$ errors. The horizontal dashed line represents the
density of field stars.
\label{fig2}}
\end{figure}

\begin{figure}
\vspace*{-10.5cm}
\includegraphics[angle=0,scale=1.0]{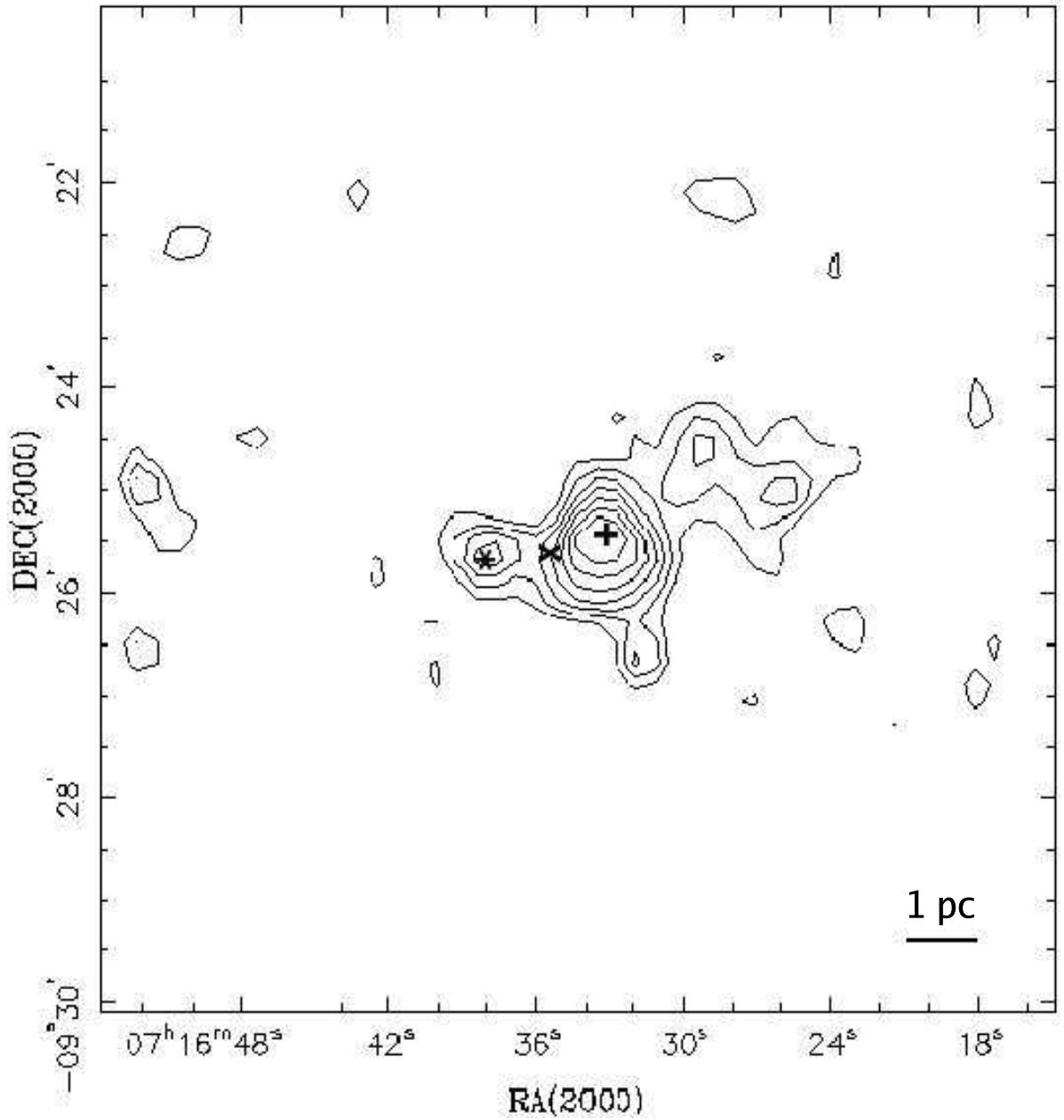}
\vspace*{-5.5cm}
\caption{The contour map of the stellar surface number density
obtained by counting stars in a 12$^{\prime\prime}\times12^{\prime\prime}$ 
($\sim 0.28~pc \times 0.28~pc$) grid for the cluster associated with 
Sh2-294 region. The contours are drawn from 6 to 22 stars/pc$^{2}$ in steps 
of 2 stars/pc$^2$. The lowest contour is at 3 times of the
 background level. The positions of the MSX and IRAS point sources are marked with asterisk
and cross symbol, respectively. The plus sign represents the possible
exciting star of the region.
\label{fig3}}
\end{figure}

\begin{figure}
\epsscale{1}
\plotone{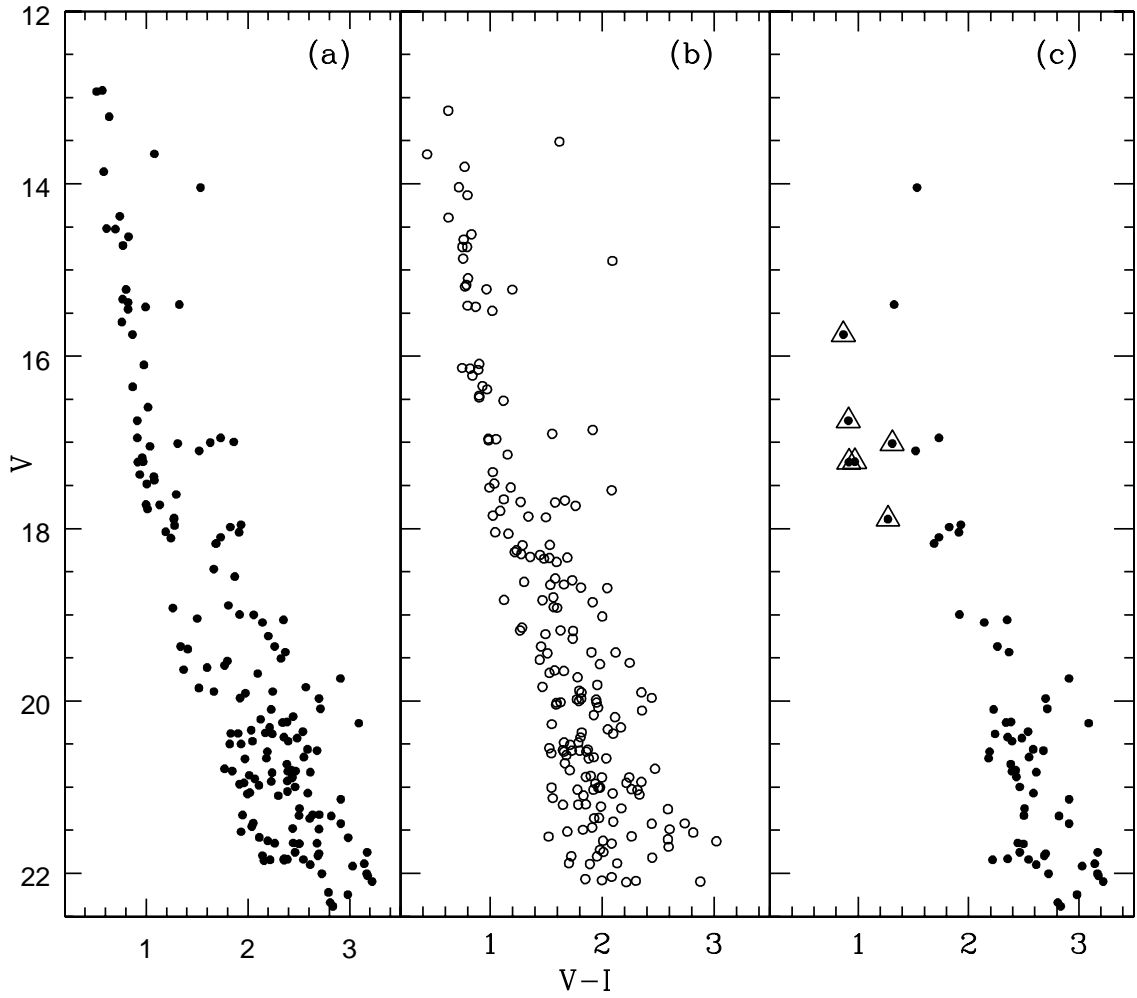}
\vspace*{-3.5cm }
\vskip.5cm
\caption{$V /V - I$ CM diagrams for (a) stars in the cluster region (b) stars
in the nearby field region of same area as of cluster region and (c) 
statistically cleaned CM diagram, where the points with triangles
 are the probable field stars (see \S 4.4).
\label{fig4}}
\end{figure}

\begin{figure}
\epsscale{1}
\plotone{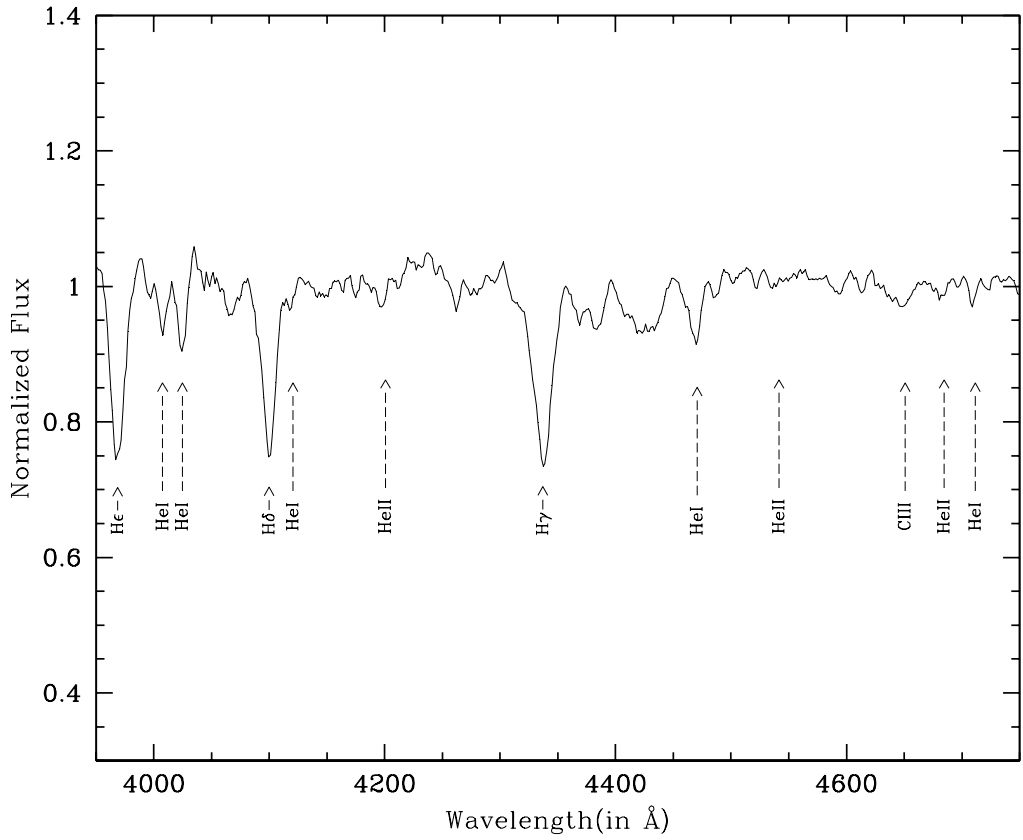}
\vspace*{-3.0cm}
\caption{The blue part of the spectrum for the ionizing source of Sh2-294 region.
The spectrum displays the characteristic similar to B0 $\pm$ 0.5 V  MS member.
\label{fig5}}
\end{figure}

\begin{figure}
\epsscale{1}
\plotone{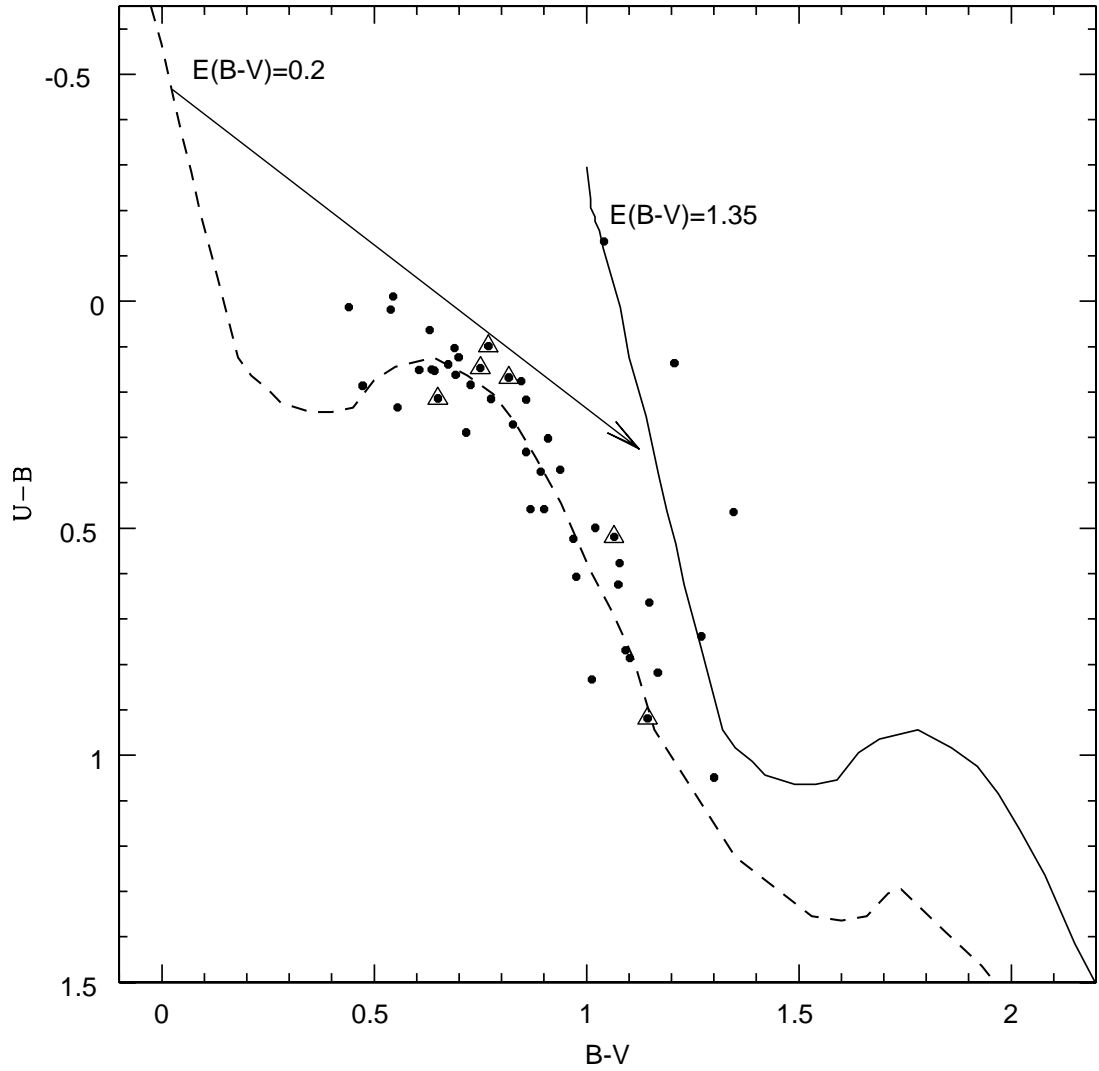}
\vspace*{-3.0cm}
\caption{Color-Color $(U - B/B - V)$ diagram for the stars lying within cluster
radius. The dashed and continuous curves represent intrinsic MS for Z=0.02
by Schmidt-Kaler (1982) shifted along the reddening vector of slope 0.72 
 (arrow line) for
$E(B - V)_{mean}$ = 0.2 mag and $E(B - V)_{min}$ = 1.35 mag,
 respectively. The 
points with triangles are likely to be probable field members identified in 
statistically cleaned CM diagram (see Fig. 4$c$).  
\label{fig6}}
\end{figure}

\begin{figure}
\epsscale{1}
\plotone{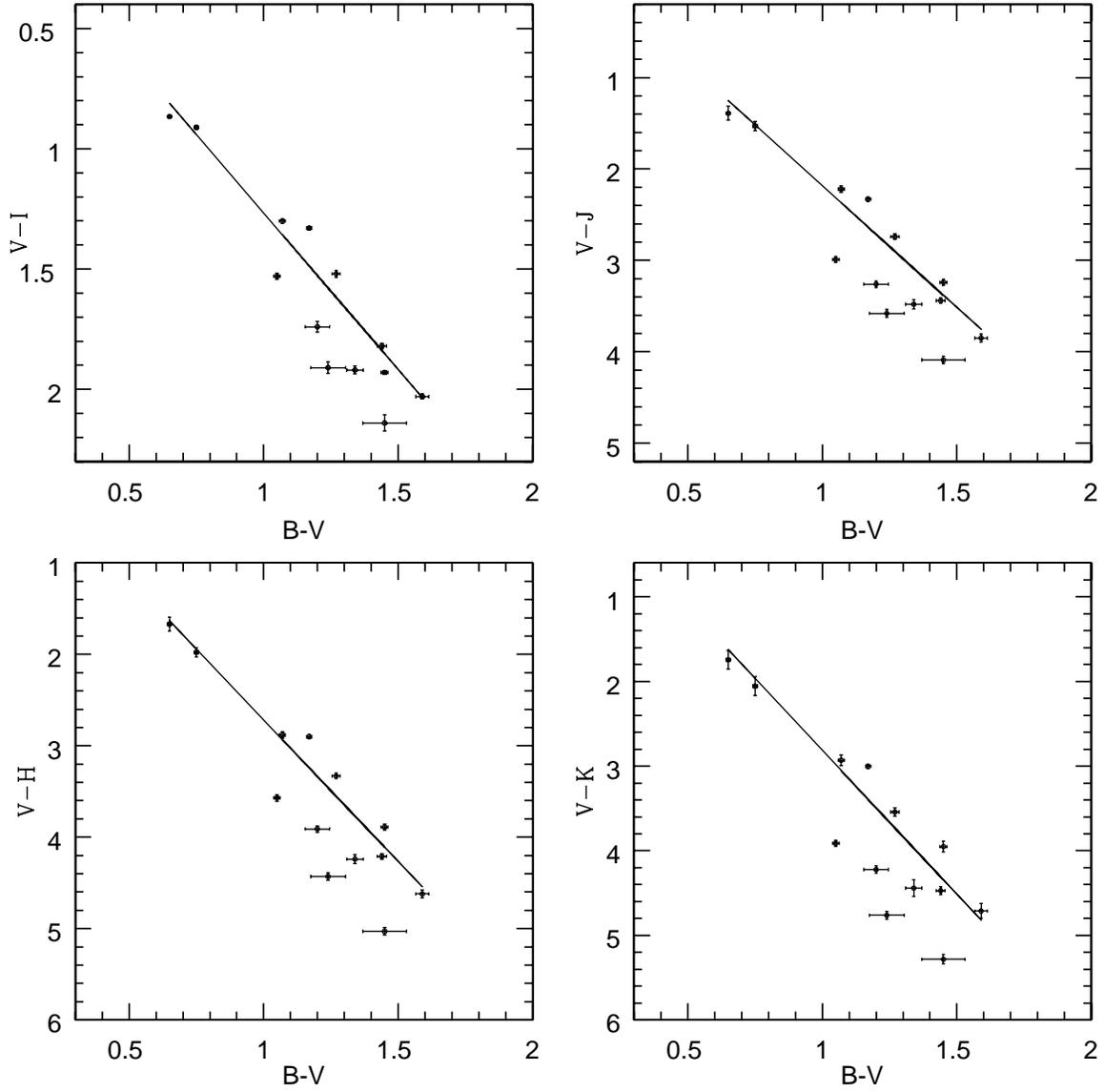}
\vspace*{-2.5cm}
\caption{$ (V - I), (V - J), (V - H), (V - K)$ vs. $(B - V)$ two color
diagrams
for the probable cluster members found in statistically cleaned CM diagram.
 Straight line shows the weighted least square fit to the data.
\label{fig7}}
\end{figure}

\begin{figure}
\epsscale{1}
\plotone{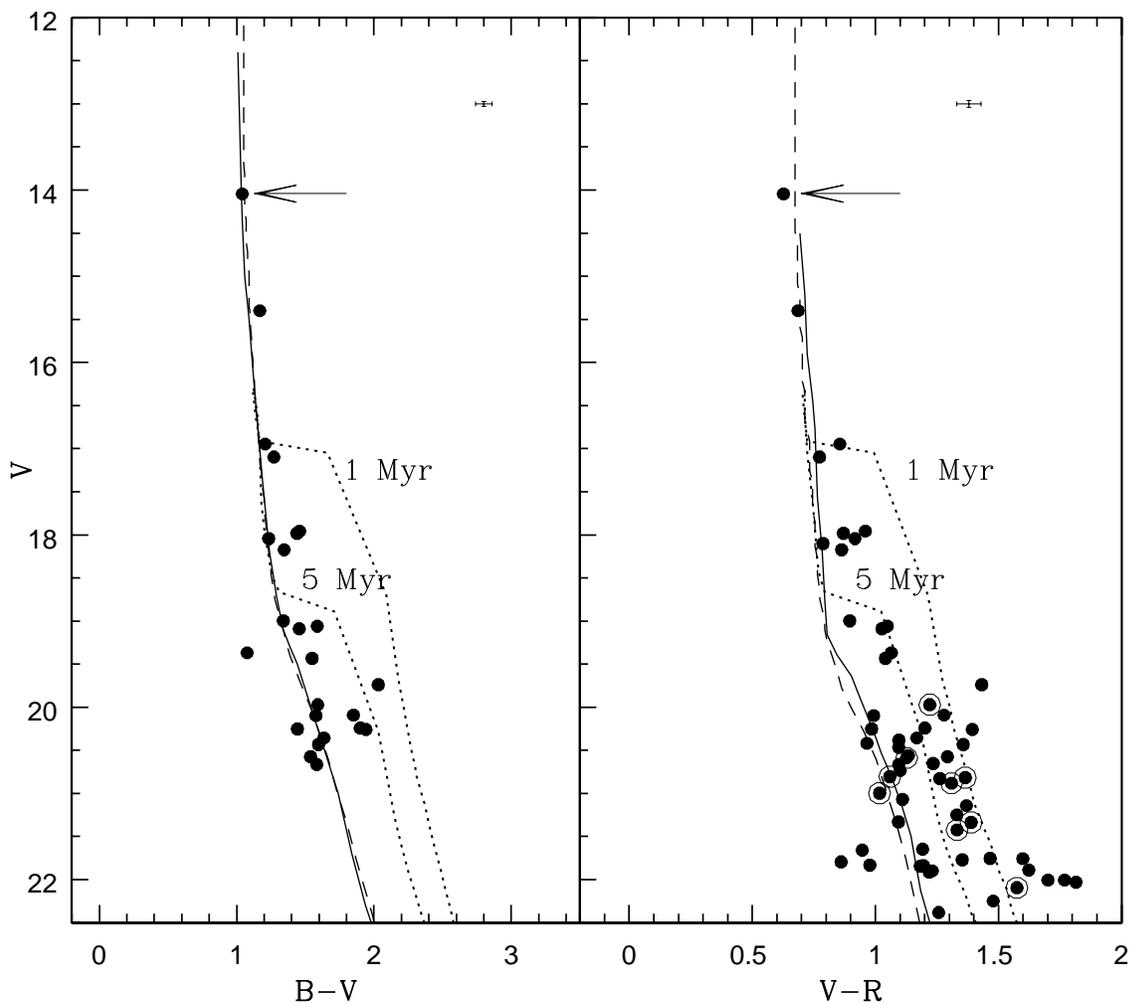}
\vspace*{-4.0cm}
\caption{Field star decontaminated V, B-V and V, V-R CM diagrams for the stars
within the cluster region. The probable field members identified
in Fig. 4$c$ and 6 are not included in these figures. 
The ZAMS in V, B-V and  V, V-R CM diagrams (solid curve)
by Schmidt-Kaler (1982) and Walker (1985) respectively, the isochrone 
of  4 Myr
(dashed curve) by Bertelli et al. (1994) and the PMS isochrones for 1 \& 5 Myr 
(dotted curve) by Siess, Dufour, \& Forestini (2000) are plotted for 
$E(B-V)_{min}$ = 1.35 mag and distance modulus of 17.6 mag. The open circles
with dots inside represent the optical counterpart of NIR excess stars
(see \S 4.6.1). The arrow points to the possible ionizing source of Sh2-294 
region. The average photometric errors are shown
in the upper right corner of the diagrams.
\label{fig8}}
\end{figure}

\begin{figure}
\epsscale{1}
\plottwo{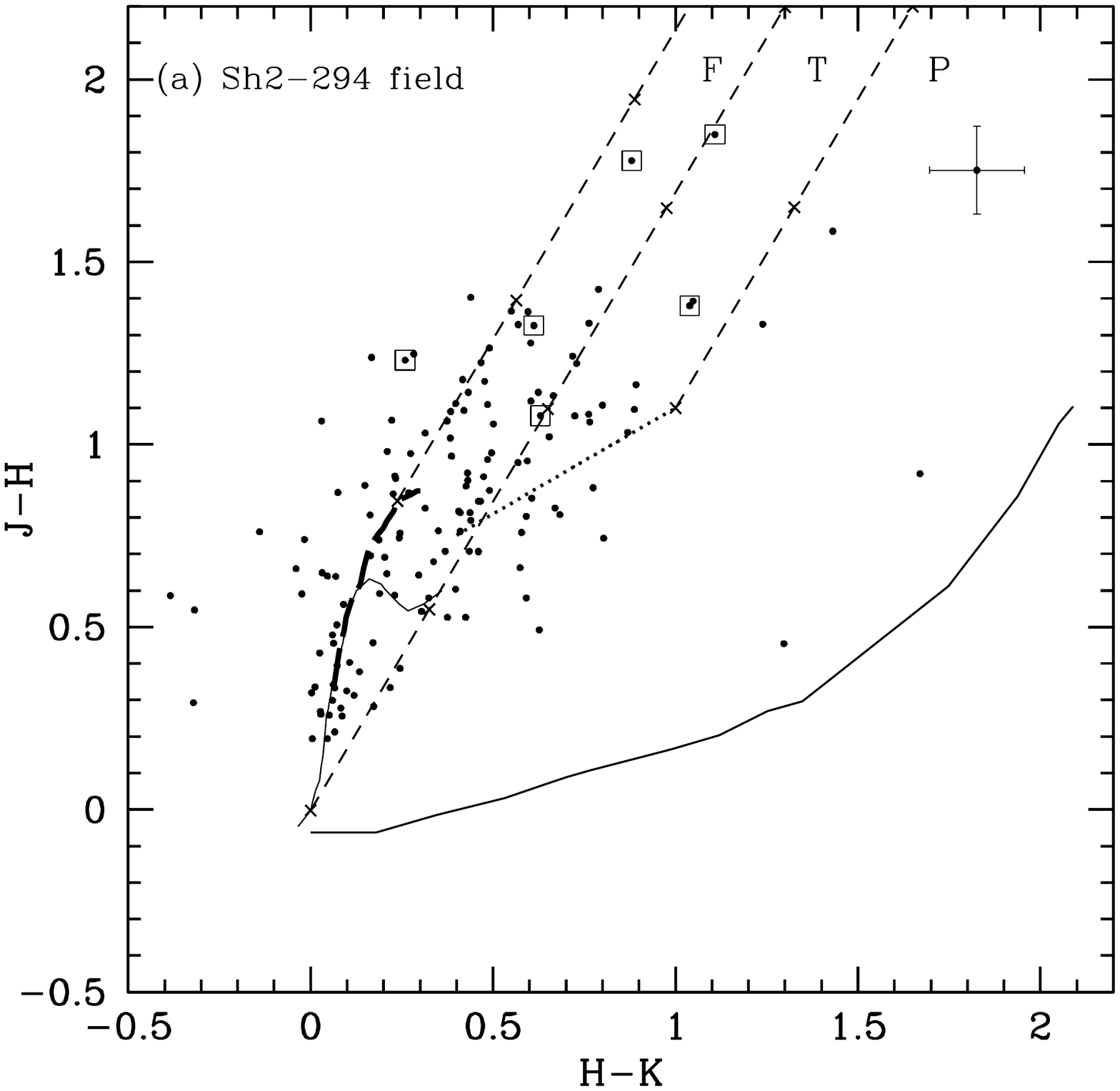}{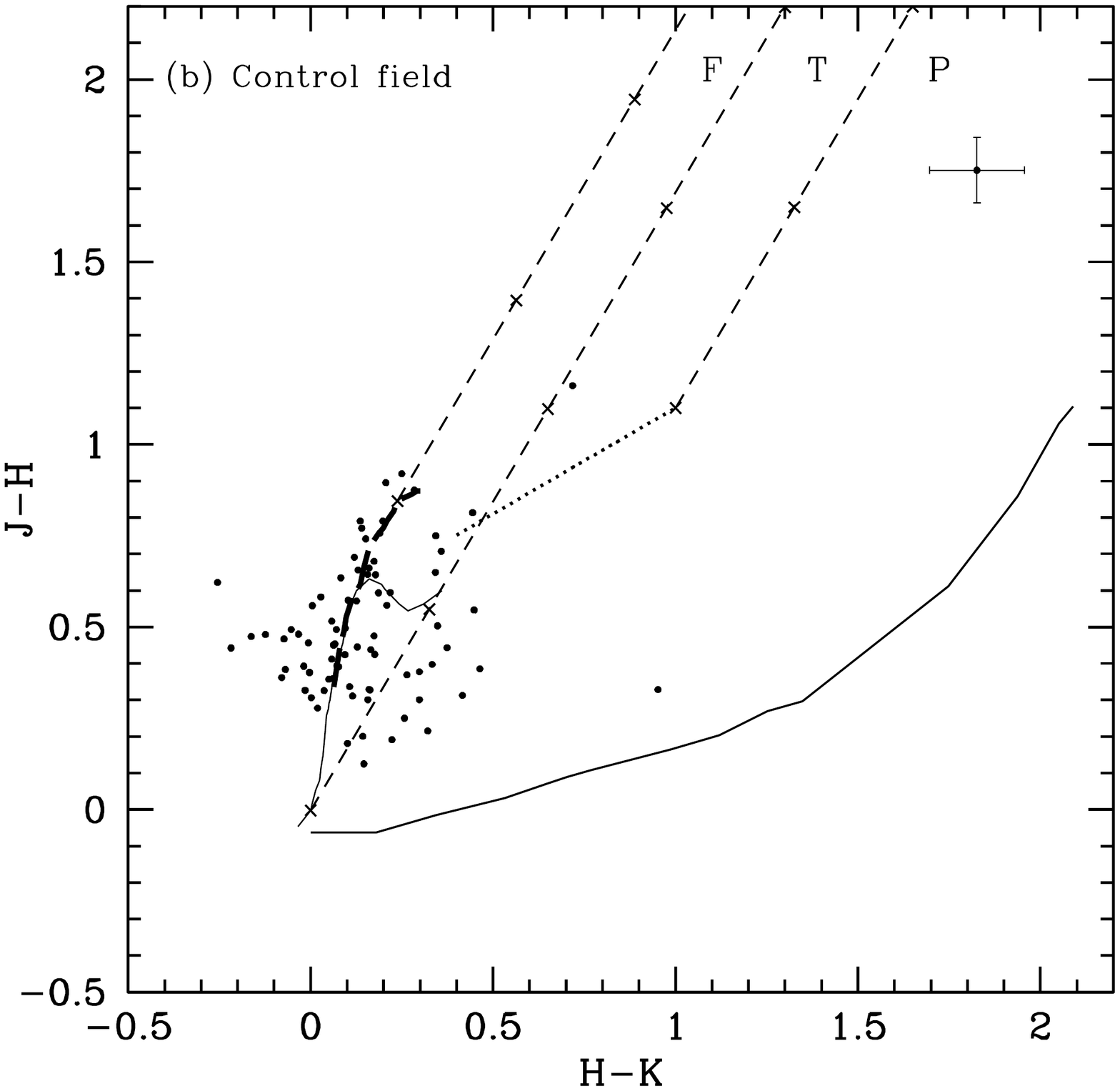}
\caption{$(H-K)$ vs. $(J-H)$  color-color diagram (a) for the cluster region in Sh2-294
and (b) for the control region. The locus of the main sequence (thin solid curve) and
giant branch (thick dashed curve) are from Bessell \& Brett (1988). The dotted line
represents locus of classical T Tauri stars from Meyer, Calvet, \& Hillenbrand (1997).
The parallel dashed lines are the reddening vectors, with crosses along
the lines corresponding to visual extinction of 5 mag. The thick continuous curve
represents the locus of Herbig Ae/Be stars (Lada \& Adams 1992). The open 
squares with dots
represent the sources located in `region A'. The plot is
classified into three regions namely ``F'',``T'' and ``P'' for different
classes of sources (see text for details). The average photometric errors are shown
in the upper right corner of the diagrams.
\label{fig9}}
\end{figure}

\begin{figure}
\epsscale{1}
\plotone{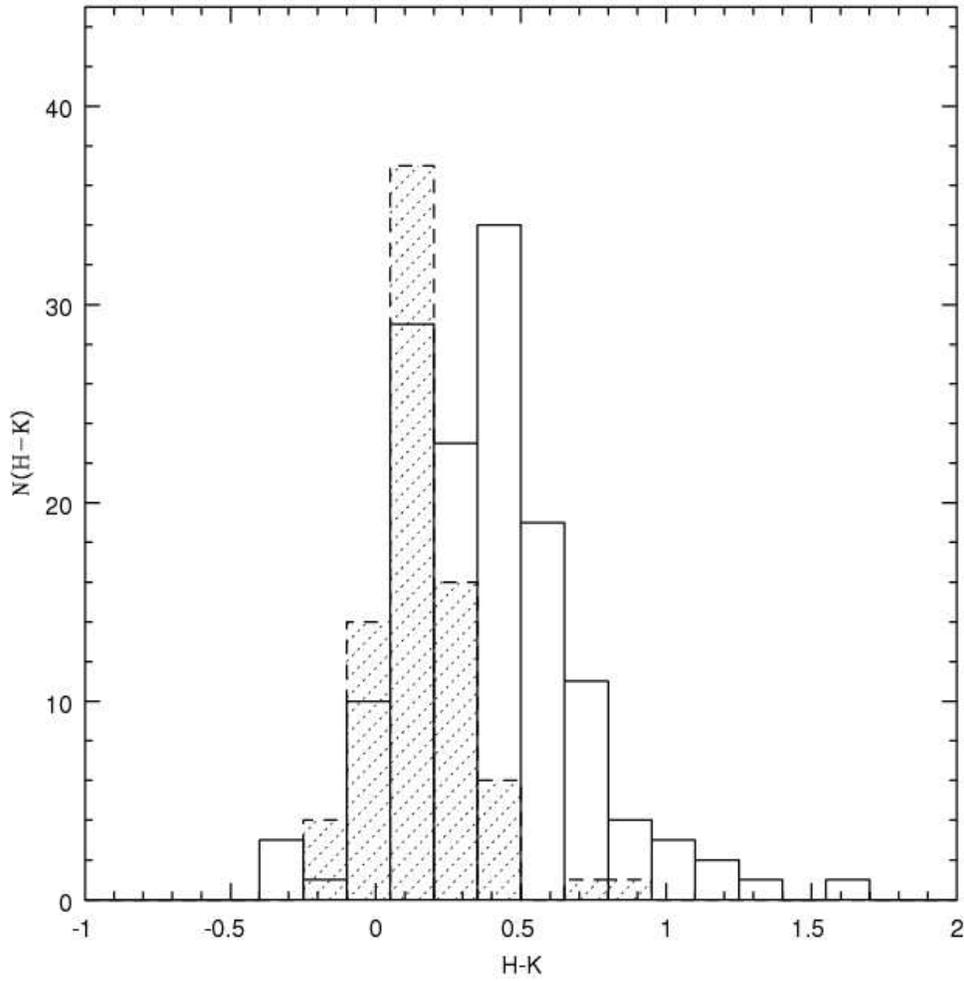}
\vspace*{-1cm}
\caption{Frequency distribution of $H-K$  colors  of the sources in the control
field (shaded area) and the Sh2-294 star forming region (thick solid line).
\label{fig10}}
\end{figure}

\begin{figure}
\epsscale{1}
\plotone{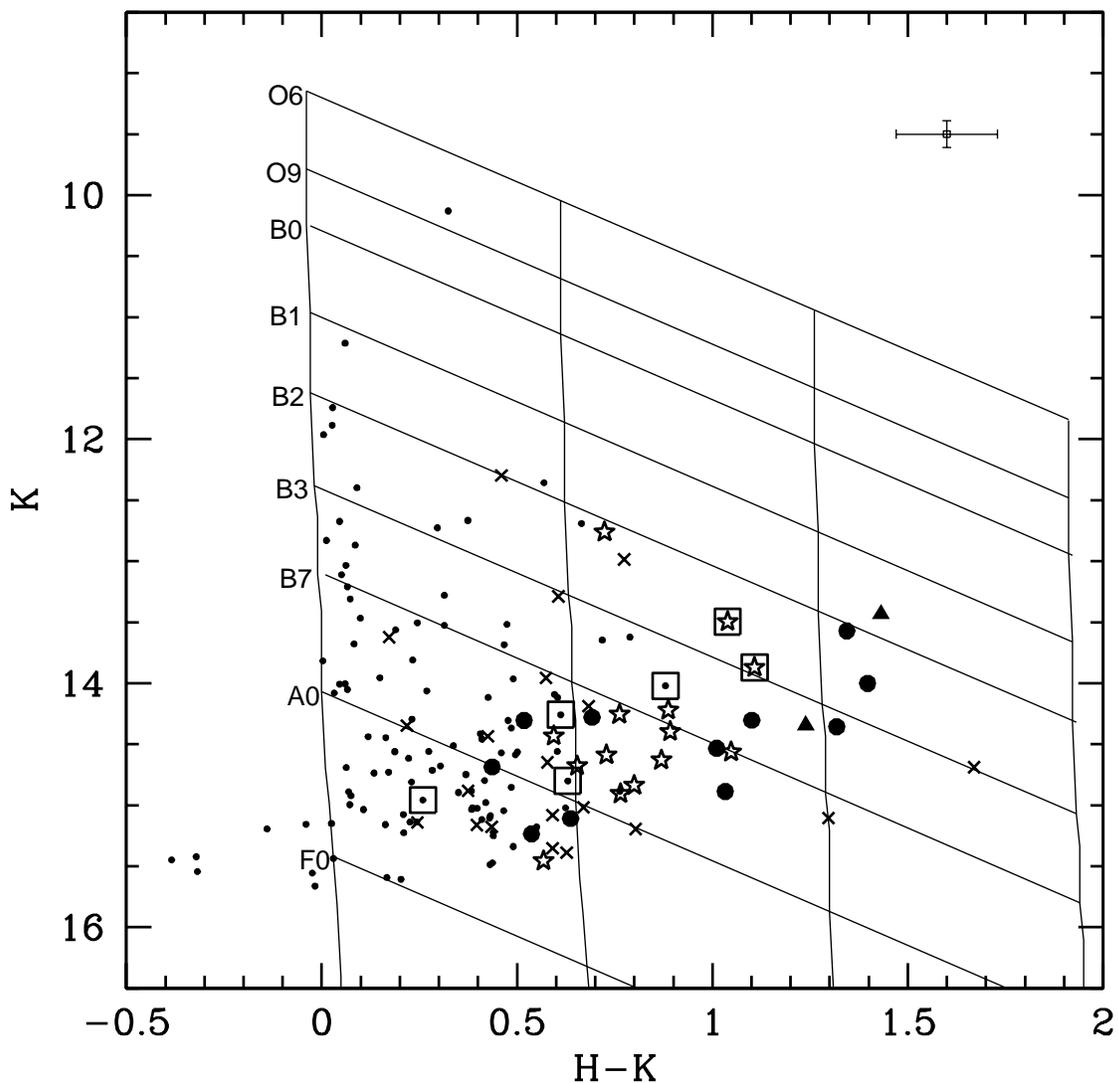}
\vspace*{-4.5cm}
\caption{Color-magnitude diagram ($(H-K)$ vs. K ) for sources detected in all the three
2MASS bands along with those detected only in $H$ and $K_{\rm s}$ bands within the cluster radius.
The thick vertical solid lines represent the ZAMS curves reddened by A$_V$=0, 10,
20 mag, respectively. The slanting lines trace the reddening vectors
for each spectral type. Star symbols represent sources of T Tauri type
(Class II), filled  triangles represent protostellar (Class I) objects, crosses
represent the probable Herbig Ae/Be stars, filled circles represent
 those sources detected in $H$ and $K_{\rm s}$ band only and the  squares
represent  the sources which lie in `region A'. The average photometric 
errors are shown in the upper right corner of the diagram.
\label{fig11}}
\end{figure}

\begin{figure}
\epsscale{1}
\plotone{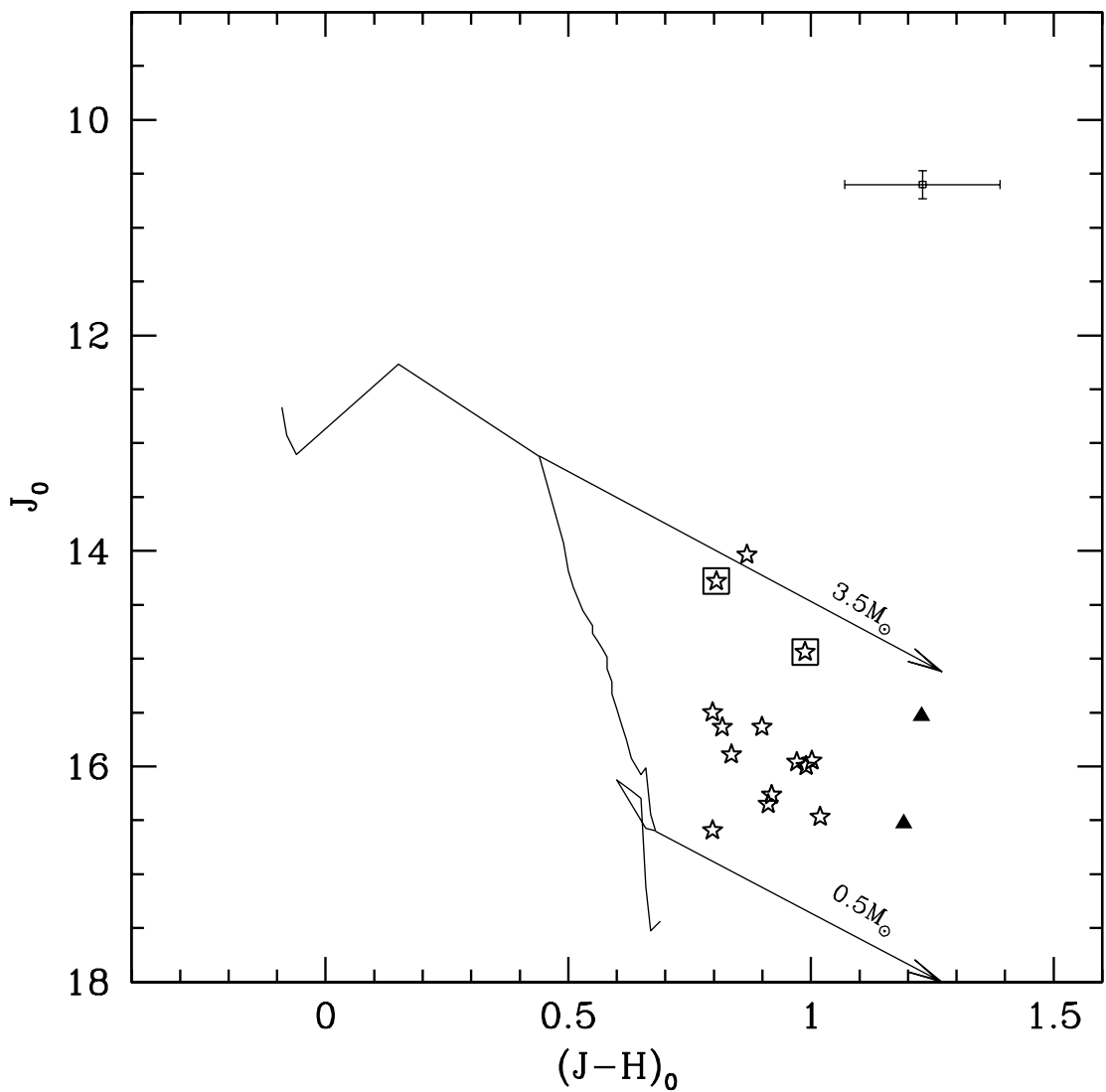}
\vspace*{-3.5cm }
\caption{ $(J-H)$ vs. $J$ intrinsic CM diagram for the YSOs (Class II and I) in Sh2-294 region. Stars are
dereddened according to their individual extinction (see text for details). The thin
solid curve represents PMS isochrones of 1$\times10^6$ yr by Siess, Dufour, \& Forestini 
(2000). The
arrows represent the reddening vectors for 0.5 and 3.5 M$_\odot$ PMS stars,
respectively. The symbols are same as in Figure 11. The average
 photometric errors of the YSOs are shown in the upper right corner of 
the diagram.
\label{fig12}}
\end{figure}

\begin{figure}
\epsscale{1}
\plotone{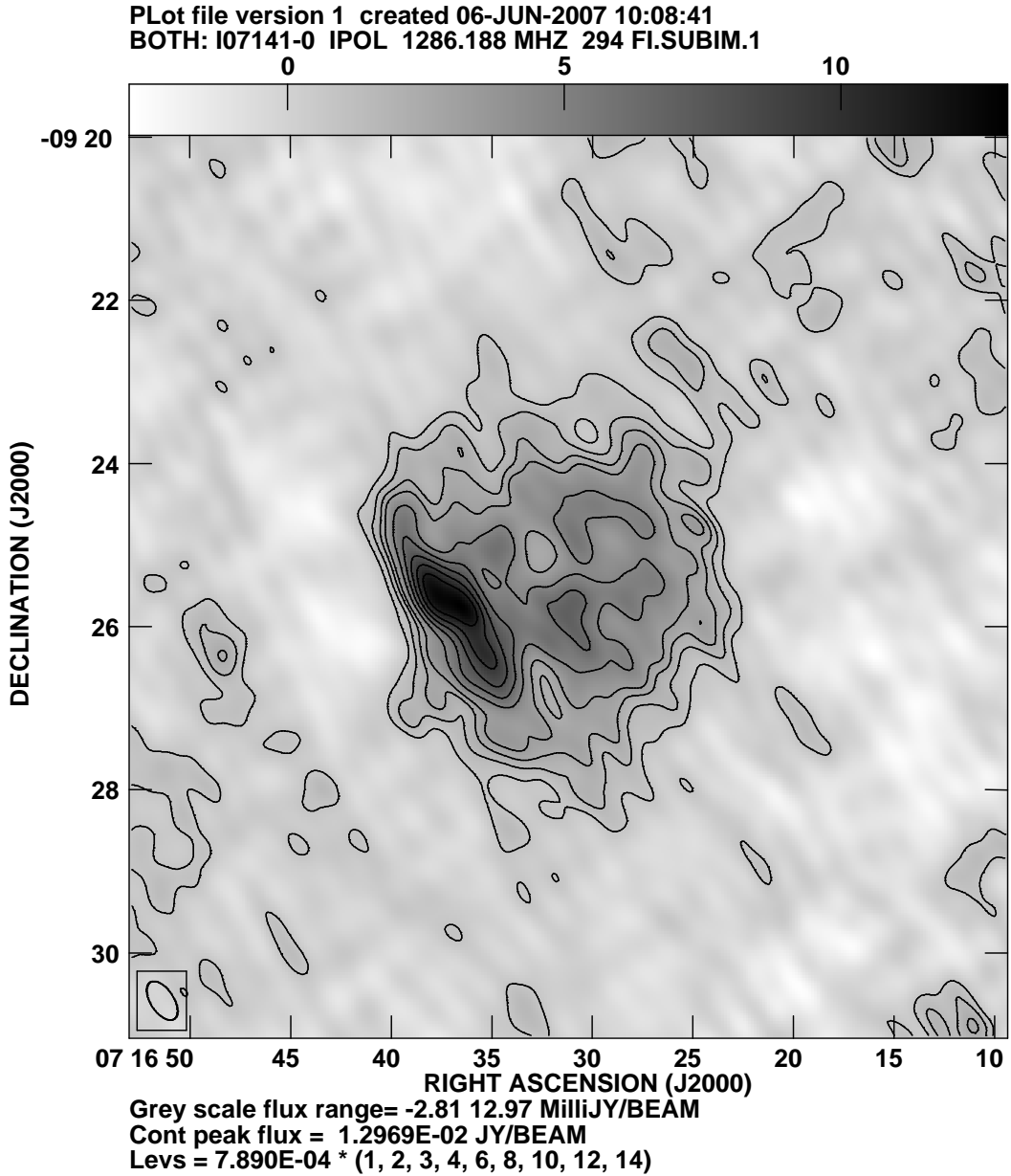}
\vspace*{-2.5cm }
\caption{GMRT low resolution map of Sh2-294 region at 1280 MHz. The resolution
is 33$^{\prime\prime}\times17^{\prime\prime}$ along PA = 33.3 deg. The contour
levels are drawn above three times of rms noise of the map, which is
0.26 mJy/beam.
\label{fig13}}
\end{figure}

\begin{figure}
\epsscale{1}
\plotone{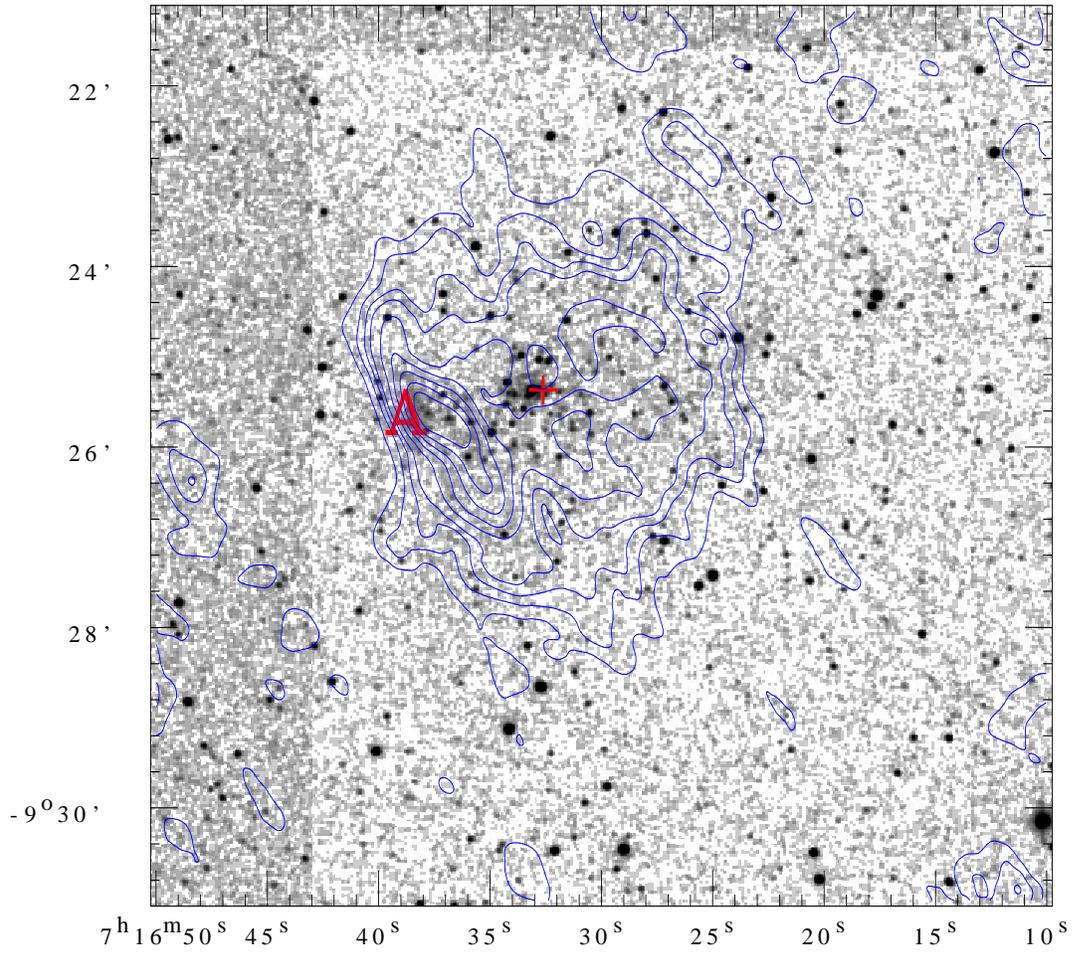}
\vspace*{-2.5cm }
\caption{2MASS $K_{\rm s}$-band image in logarithmic scale overlaid by the GMRT radio continuum contours at
1280 MHz (blue contours). The plus sign marks the position of the exciting source of 
Sh2-294 region. The `region A' is shown in the image.
\label{fig14}}
\end{figure}

\clearpage

\begin{figure}
\includegraphics[angle=0,scale=0.8]{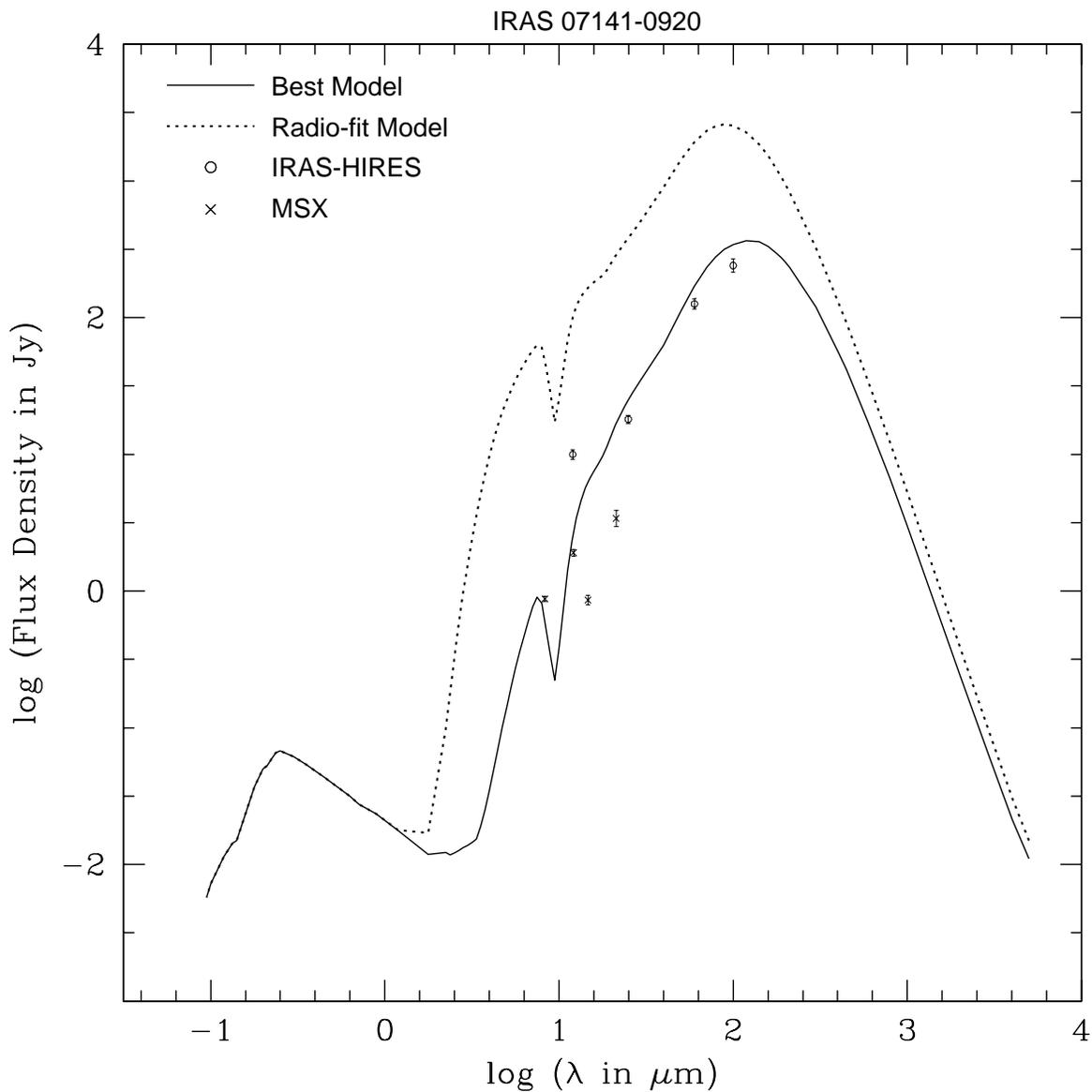}
\vspace*{-4cm }
\caption{Comparison of the SED from observations and the radiative 
transfer models of the interstellar cloud near `region A'. The open circles 
denote $IRAS-HIRES$ data, while the crosses represent the $MSX$ data.
The $HIRES$ fluxes are integrated over 3\arcmin~diameter circle centred on $IRAS$ 
peak. The solid curve denotes our best-fit model to the data and dashed 
curve represents the radio-fit model (see the text).  
\label{fig15}}
\end{figure}

\begin{figure}
\epsscale{1}
\plotone{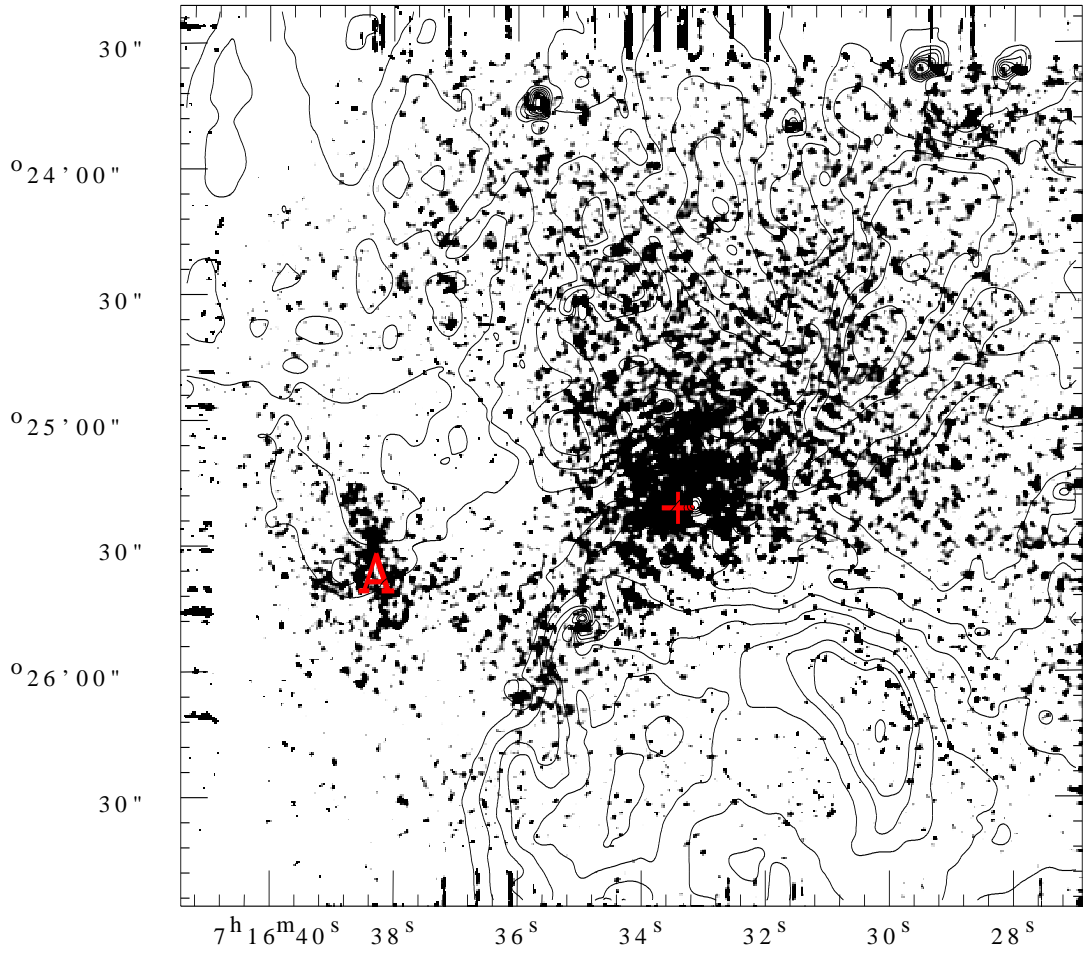}
\vspace*{-3.0cm}
\caption{The H$\alpha$ continuum-subtracted contours overlaid over the
continuum-subtracted H$_2$ (1-0)S1 line image. The prominent peaks
in the contours are due to the residuals of continuum subtraction around some 
stars (see Fig. 17). The symobls are same as in Fig. 14.
\label{fig16}}
\end{figure}

\begin{figure}
\epsscale{1}
\plotone{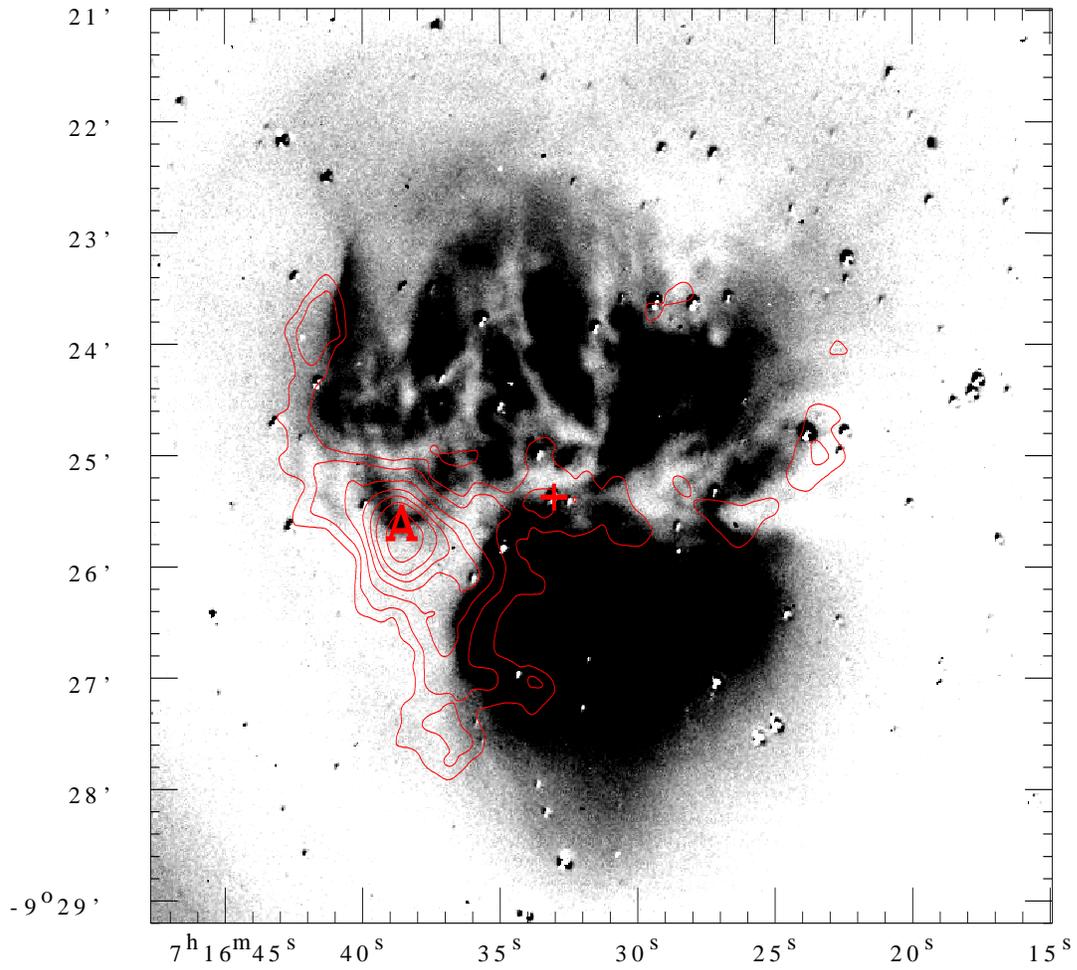}
\vspace*{-3cm}
\caption{Contour diagram from the MSX A-band image (8.28 $\mu$m), overlaid
on an H$\alpha$ line image. The contours start at 2.1$\times10^6$
W m$^{-2}$ Sr$^{-1}$ and the innermost contour corresponds to a value of
2.7$\times10^6$ W m$^{-2}$ Sr$^{-1}$. Residuals of the continuum subtraction
 around some of the stars are seen in the image. The plus sign represents 
the position of the
central ionizing star that emits UV photons.
\label{fig17}}
\end{figure}

\begin{figure}
\epsscale{1}
\plottwo{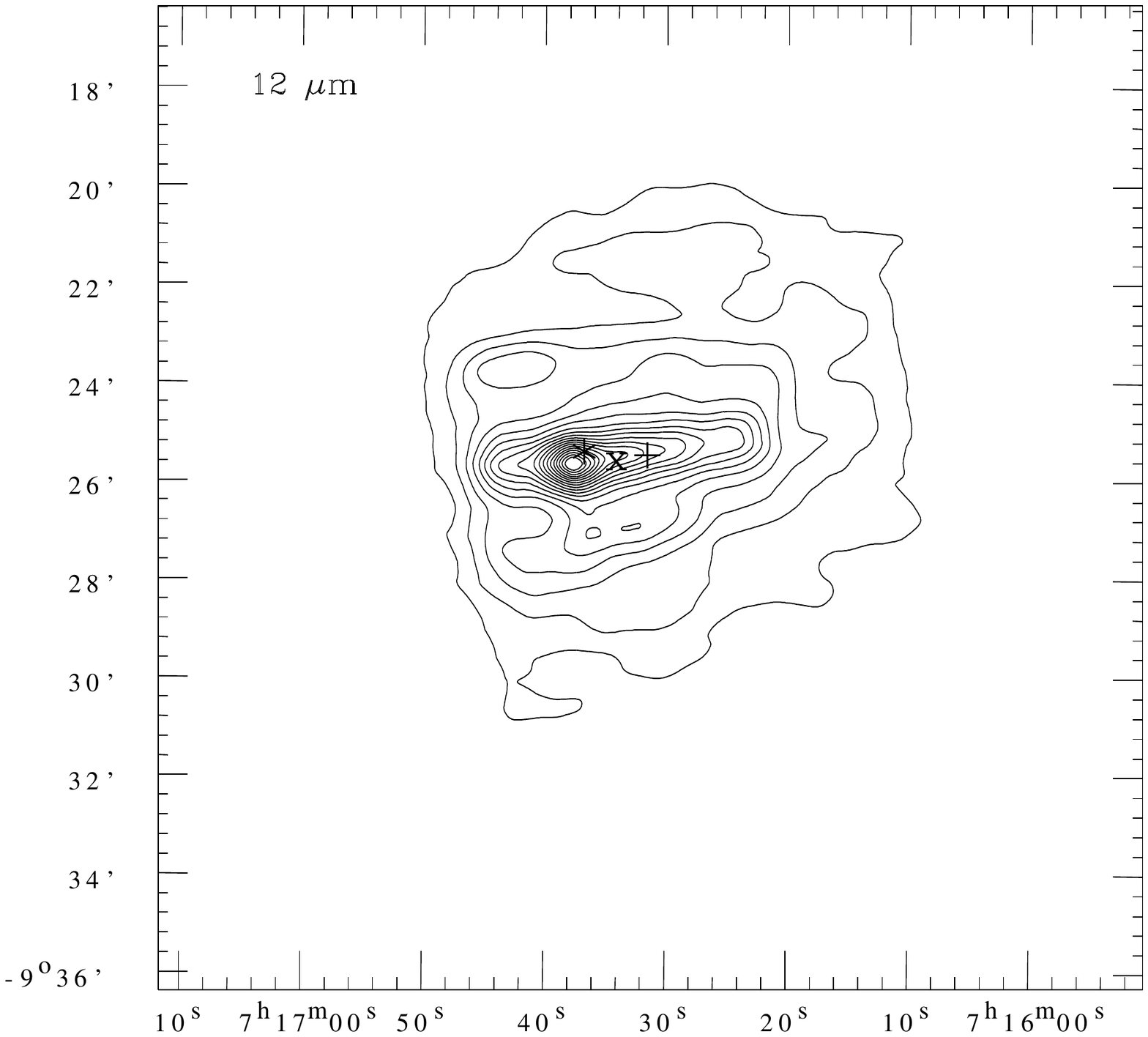}{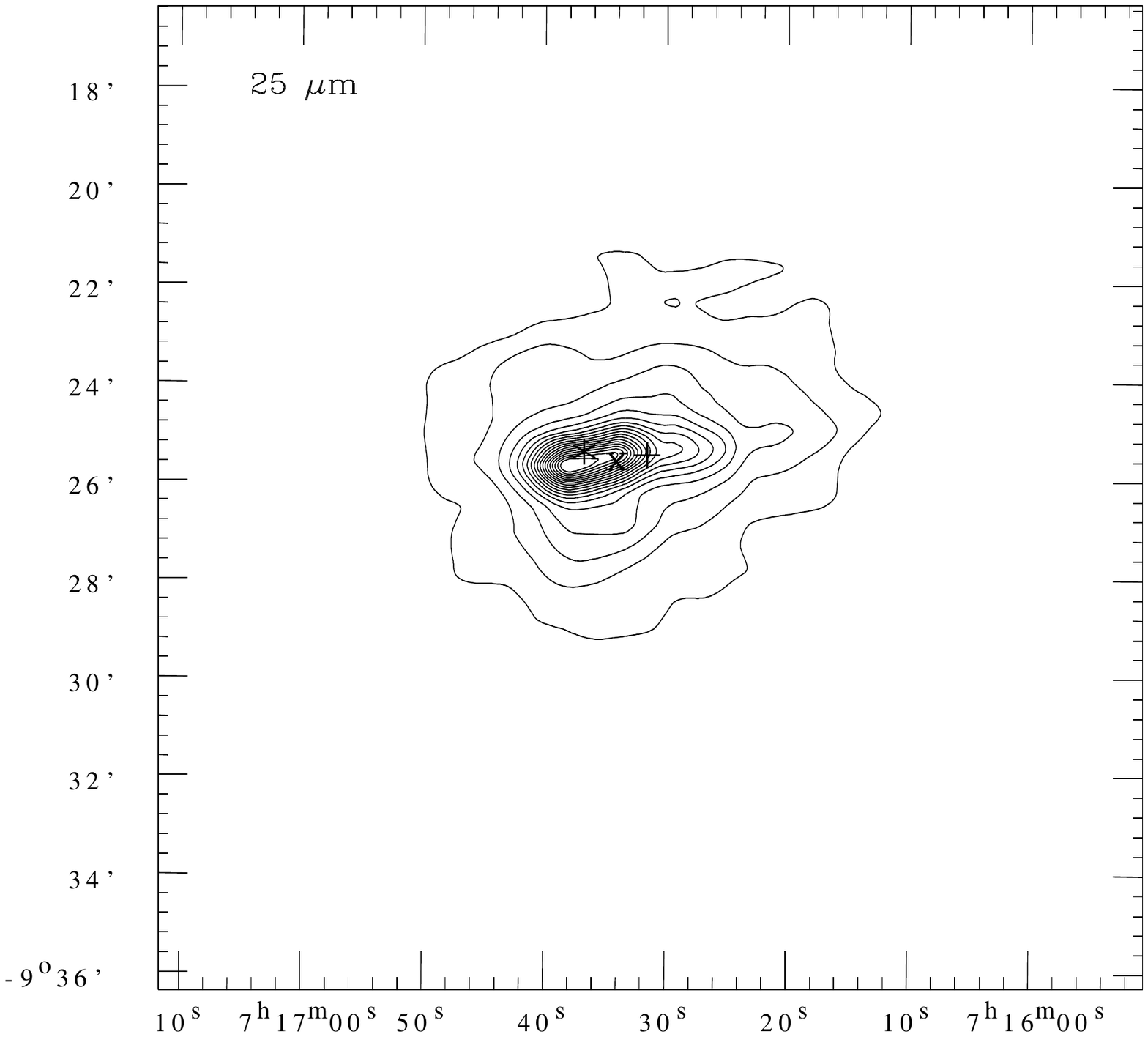}

\vspace*{-1.5cm}

\plottwo{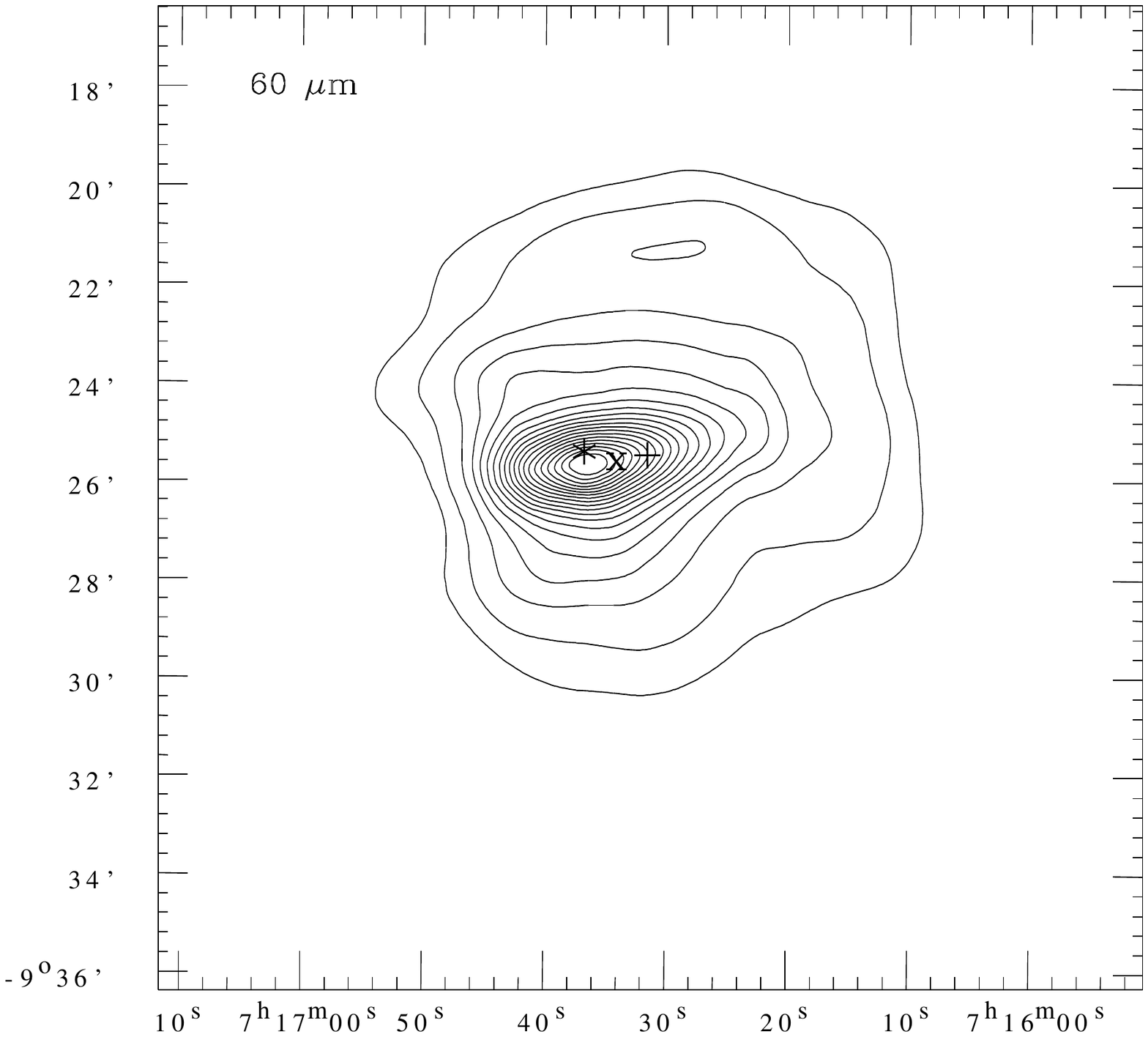}{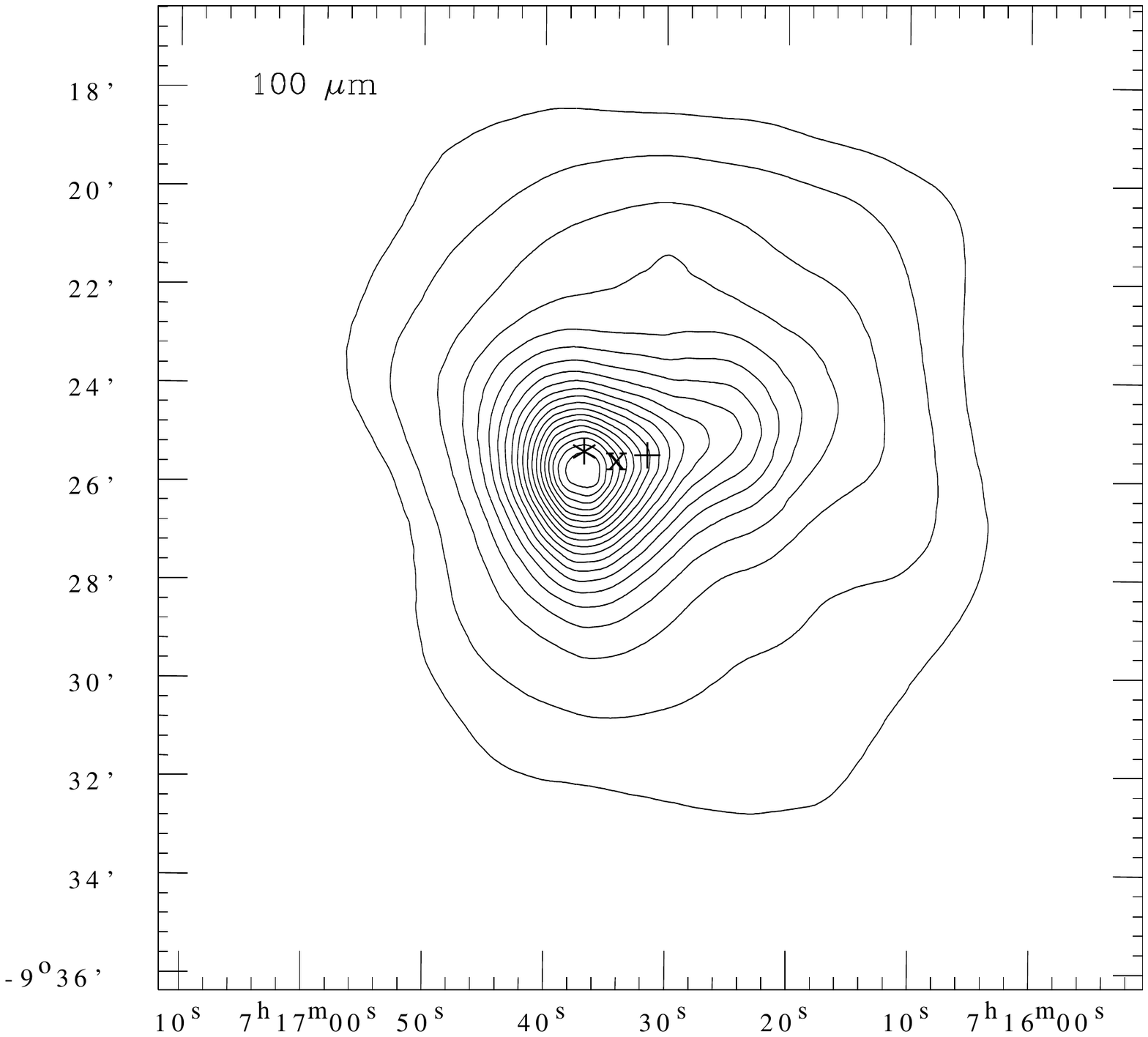}
\vspace*{-1cm}
\caption{HIRES processed IRAS maps for Sh2-294 region in the four bands (clockwise
from the top left)-{\it Top left)} 12 $\mu$m with peak = 41 MJy/Sr, {\it Top right)}
25 $\mu$m with peak = 80.3 MJy/Sr, {\it Bottom left)} 60 $\mu$m with peak = 384.2 MJy/Sr,
and {\it Bottom right)} 100 $\mu$m with peak = 600.5 MJy/Sr. The isophot contour levels
in 12 \& 25 $\mu$m are 95, 90, 80, 70, 60, 50, 40, 30, 20, 10 \& 5\%, and in 60 \& 100 $\mu$m
are 95, 90, 80, 70, 60, 50, 40, 30, 20, 10, 5 \& 2.5 \% of the respective peaks.
The abscissa and the ordinates are in J2000.0 epoch. The symbols are same as in Fig. 3.
\label{fig18}}
\end{figure}

\begin{figure}
\epsscale{1}
\plottwo{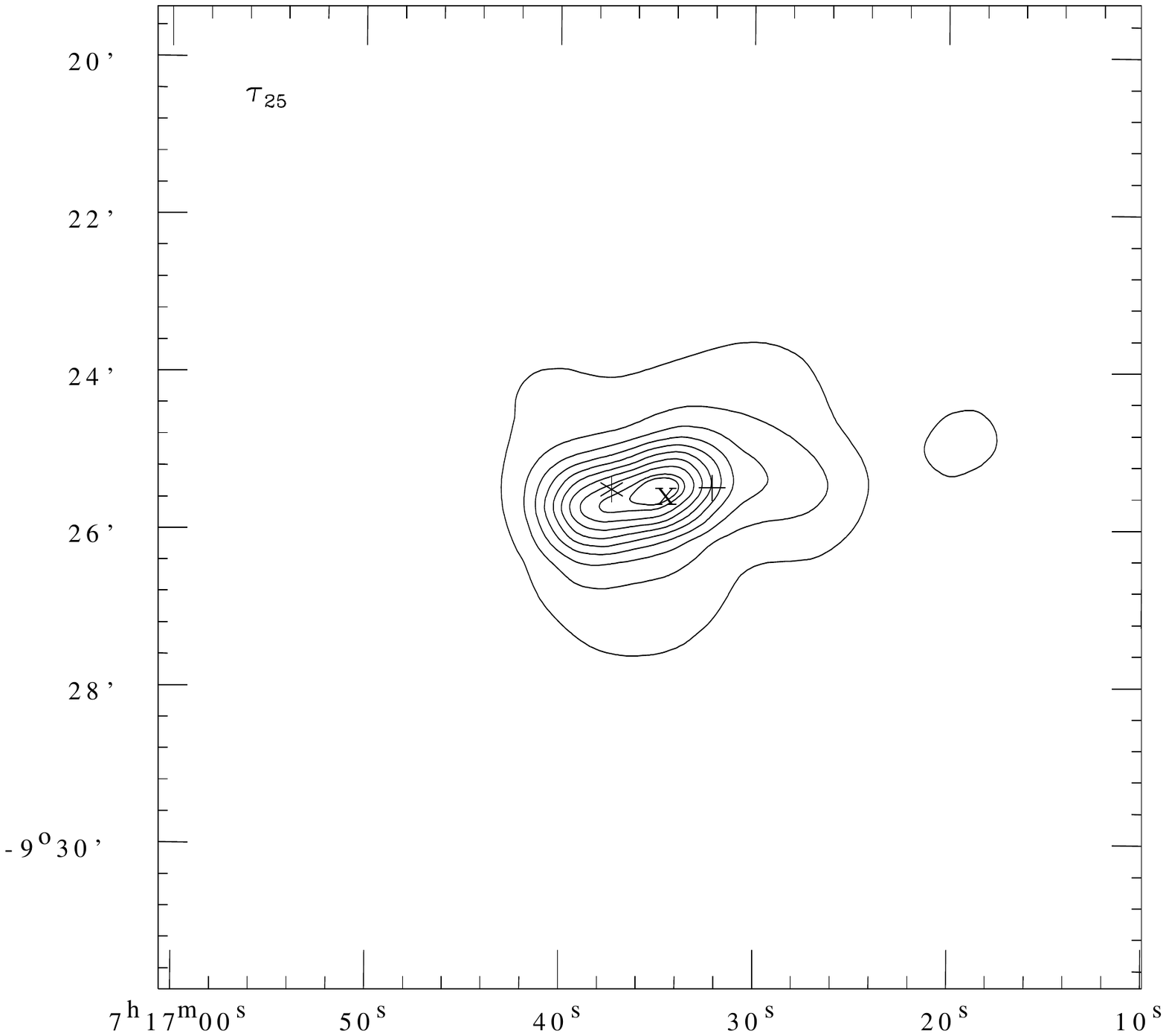}{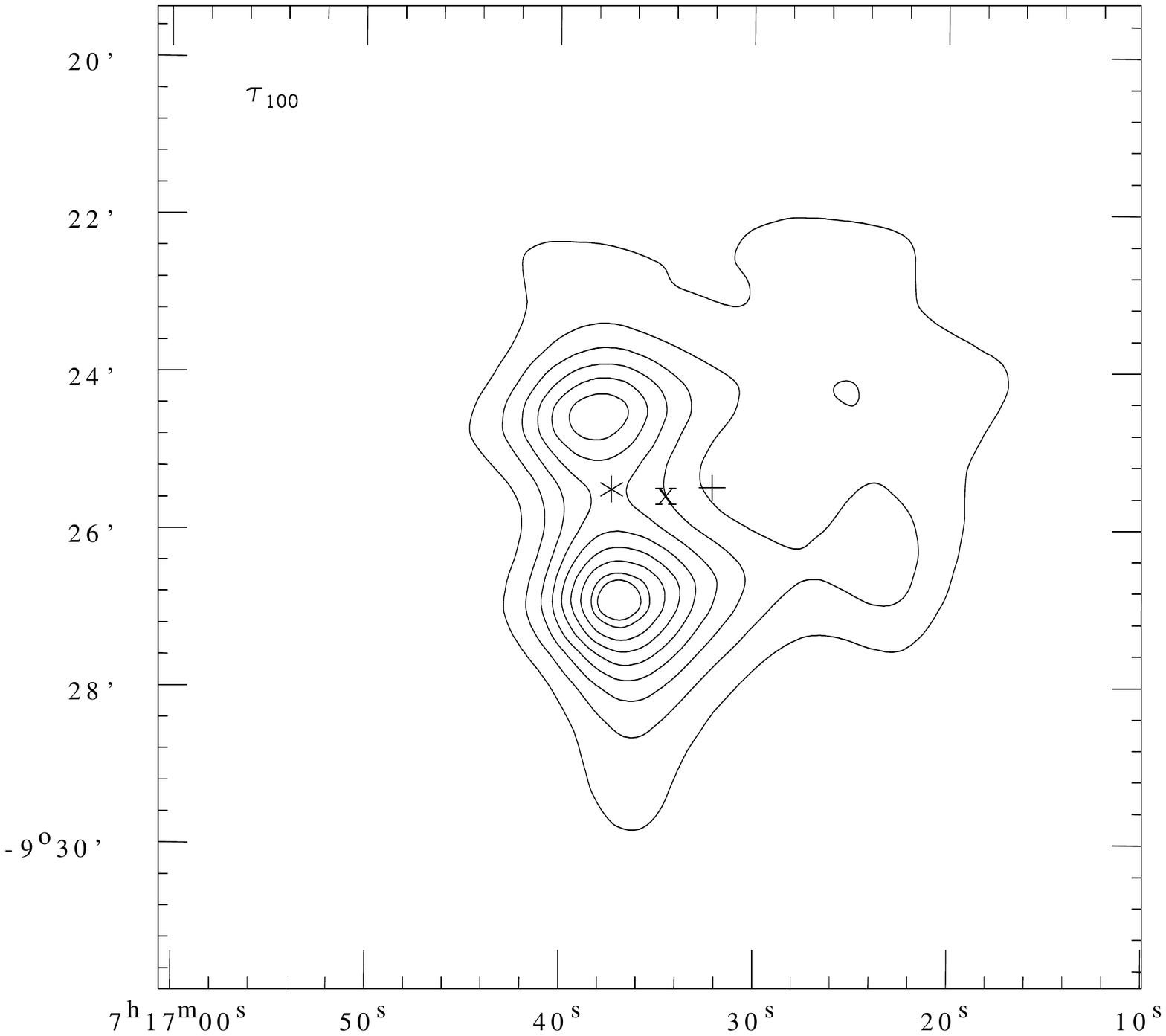}
\vspace*{-0.5cm}
\caption{$Left:$ The dust optical depth ($\tau_{25}$) distribution from the HIRES 12 and
25 $\mu$m maps for the region around Sh2-294 assuming a dust emissivity law of
$\epsilon_{\lambda} \propto \lambda^{-1}$. The contours represent 95, 90, 80, 70, 60, 50,
40, 30, 20 \& 10\% of the global peak value of 7.9$\times$10$^{-7}$. $Right:$ The dust
optical depth ($\tau_{100}$) distribution from the HIRES 60 and 100 $\mu$m maps
for the region around Sh2-294 assuming a dust emissivity law of
$\epsilon_{\lambda} \propto \lambda^{-1}$. The contours represent 95, 90, 80, 70, 60, 50,
40, 30, 20 \& 10\% of the global peak value of 1.2$\times$10$^{-3}$. The symbols
are same as in Fig. 3.
\label{fig19}}
\end{figure}

\begin{figure}
\epsscale{1}
\plotone{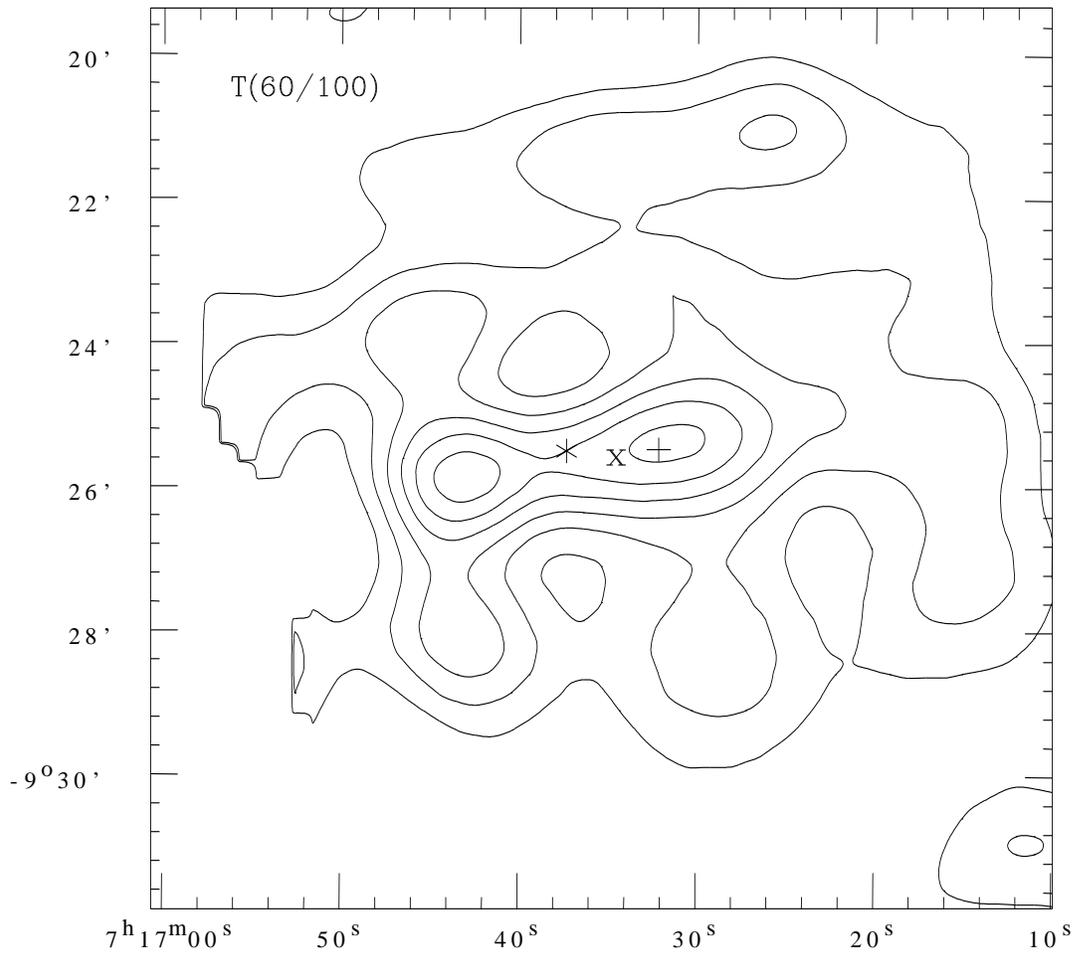}
\vspace*{-3cm}
\caption{The dust color temperature ($T(60/100)$) distribution. The contours
correspond to the temperatures from 30 to 40 K in steps of 2 K. The abscissa and
the ordinates are in J2000.0 epoch. The symbols are same as in Fig. 3.
\label{fig20}}
\end{figure}

\begin{figure}
\epsscale{1}
\plotone{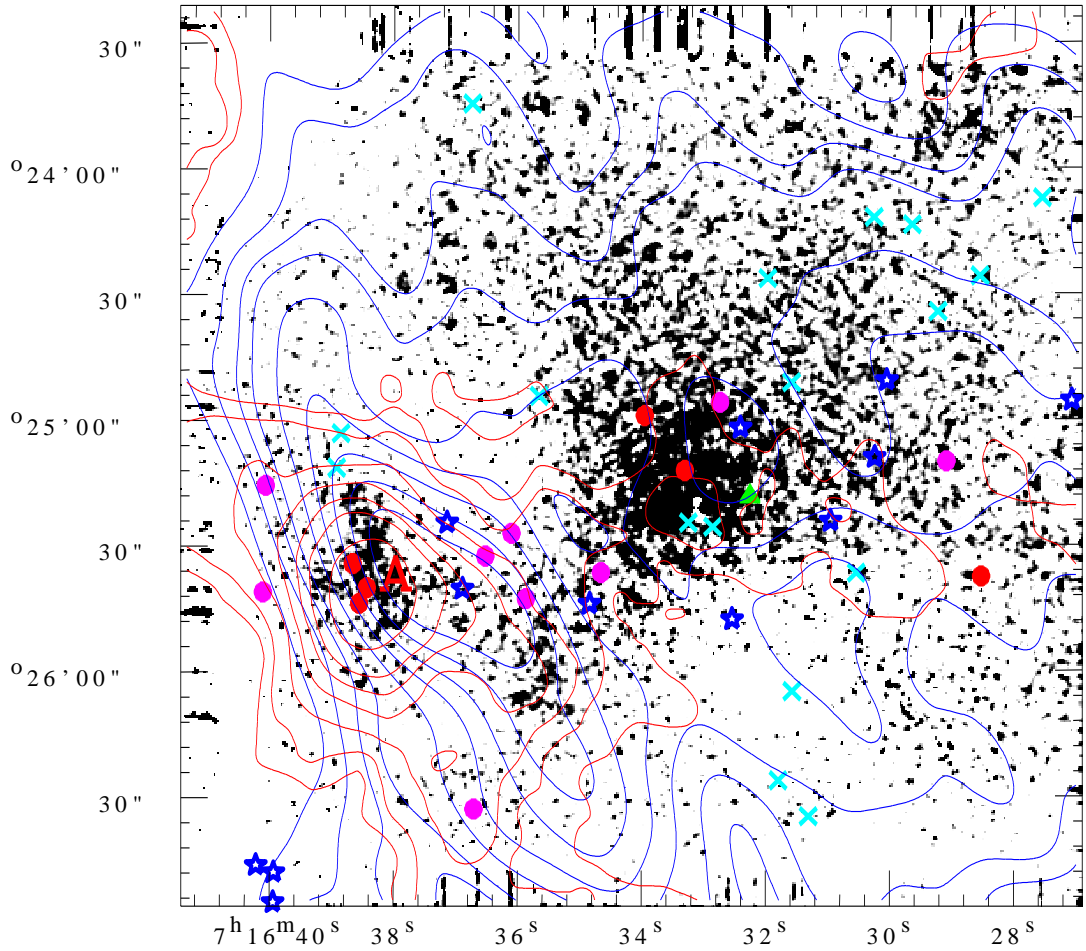}
\vspace*{-3cm}
\caption{The continuum-subtracted molecular hydrogen 2.12 $\mu$m line
image from HCT overlaid by the GMRT radio continuum contours (blue color). The
red contours represent the distribution of warm dust from MSX A-band
(8.28 $\mu$m). Spatial distributions of Class II (star symbols), Class I 
(filled  triangles), probable
Herbig Ae/Be sources (crosses), sources detected only in $H$ and $K_{\rm s}$ 
bands (filled magenta circles) and sources detected only in $K_{\rm s}$ band 
(filled red circles) are shown in the figure.
\label{fig21}}
\end{figure}

\end{document}